\documentclass[a4paper,11pt]{article}
\usepackage{jheppub}

\usepackage{amsmath, amssymb, bm,slashed, color, graphicx}
\usepackage{caption}
\usepackage{subcaption}
\usepackage{braket}
\usepackage{epsfig}
\usepackage{hyperref,enumerate}
\usepackage{pstricks}
\usepackage{slashed}
\usepackage{amsfonts}
\usepackage{url}
\usepackage{color}
\usepackage[T1]{fontenc}
\usepackage{lmodern}
\allowdisplaybreaks
\hypersetup{colorlinks,citecolor= nicegreen,linkcolor= nicered}
\definecolor{nicered}{rgb}{0.7,0.1,0.1}
\definecolor{nicegreen}{rgb}{0.1,0.5,0.1}

\newcommand{\beq}{\begin{equation}}
\newcommand{\eeq}{\end{equation}}

\usepackage[normalem]{ulem}

\def\({\left(}
\def\){\right)}
\def\[{\left[}
\def\]{\right]}

\title{\boldmath
Exploring the Top-Higgs FCNC Couplings at Polarized Linear Colliders with Top Spin Observables}

\author[a]{Bla\v zenka Meli\'c, Monalisa Patra}

\affiliation[a]{Institut Ru\dj er Bo\v skovi\'c, Theoretical Physics Division, \\Bijeni\v cka 54, HR-10000 Zagreb, Croatia}

\abstract{ 
We study the nature of flavor changing neutral couplings of the top quark with the Higgs boson
and the up/charm quark in the $t\bar{t}$ production at linear colliders. 
There are previous bounds on such $tqH$ couplings at both, linear and hadronic colliders, with the 
assumption that it couples equally to the left and the right handed fermions. 
In this paper we examine the chirality of the $tqH$ coupling and construct different observables
which will be sensitive to it. The kinematics of the emitted $q$ from $t\rightarrow qH$
in $t\bar{t}$ production is
discussed and it was found that the polar angle distribution of $q$ is sensitive to the chiral nature of $tqH$ couplings. The
observables in the context of top-antitop spin correlations,
which are sensitive to new physics in the top decay are considered using different spin-quantization  
bases. It was found that in particular the off-diagonal basis can be useful to  distinguish among the chiral $tqH$ couplings. 
The sensitivity of the unpolarized ILC in probing the couplings at 
the 3$\sigma$ level at $\sqrt{s}$ = 500 GeV and $\cal L$ = 500 fb$^{-1}$
is also studied, resulting in predicted BR$(t\rightarrow qH) < 1.19 \times 10^{-3}$.
This limit is further improved to BR$(t\rightarrow qH) < 8.84 \times 10^{-4}$ with the inclusion of initial beam polarization of left
handed electrons and right handed positrons.}

\begin{document}
\maketitle

\section{Introduction}

The search for the Flavor Changing Neutral Current (FCNC) processes, has
been one of the leading tools to test
the Standard Model (SM), in an attempt of either discovering or putting
stringent limits on new physics
scenarios. The discovery of the Higgs boson at the LHC, has lead the way to
a comprehensive program of measuring of 
its properties and branching ratios, in order to look for deviations
from the SM predicted Higgs. Within the SM, there are no FCNC transitions at tree level
mediated by the Higgs boson, due to the the presence of only one 
Higgs doublet and at the one-loop level these FCNC interactions are extremely small.  There are  however many extensions of the SM
where the suppression of the neutral flavor changing transitions due to the
Glashow-Iliopoulos-Maiani (GIM) mechanism 
can be relaxed, with the presence of additional scalar doublets or through
the additional contributions of new particles in the loop diagrams. 
In the presence of two or more scalar doublets, these FCNC interactions will be generated at tree level and can be very large unless some ad-hoc discrete symmetry is imposed.
 
Motivated by the nature of the standard Yukawa coupling scheme 
the authors of~\cite{Cheng:1987rs} observed that the new FCNC couplings in the general two-Higgs doublet model  naturally follow the hierarchical structure of the quark masses and therefore 
any $\bar{q}q' H$ coupling should experience the following structure 
\begin{equation}
g_{qq'H} \sim \sqrt{m_q m_q'},
\end{equation} 
indicating that the larger couplings can be expected in the FCNC interactions of a top-quark with the Higgs field.
The large production rate of the top quarks at the LHC allows one to look for a transition of the top quark to a quark of a different flavor but same charge, $t\rightarrow cH$ (and $t\rightarrow uH$),
as no symmetry prohibits this decay. The SM branching ratio of this process is extremely small, of the order BR($t\rightarrow cH)_{{\rm SM}} \approx 10^{-15}$~\cite{Mele:1998ag,AguilarSaavedra:2004wm}, 
which is many orders of magnitude smaller than the value to be measured
at the LHC at 14 TeV. Therefore an affirmative observation of the process $t\rightarrow qH$, well above the SM rate,
will be a conclusive indication of new physics beyond the SM. 

The probing of FCNC couplings in the quark sector, can be performed either at
a high energy collider or indirect limits can be obtained from neutral meson 
oscillations ($K^0-\bar{K}^0$, $B^0-\bar{B}^0$ 
and $D^0-\bar{D}^0$)~\cite{Bona:2007vi, Blankenburg:2012ex, Aranda:2009cd}. 
%The indirect limit on the $tqH$ coupling is obtained from the low energy
%measurement of the $D^0-\bar{D}^0$ mixing process~\cite{Aranda:2009cd}. 
The $tqH$ coupling also affects the $Z\rightarrow c\bar{c}$ decay at the loop
level and is therefore constrained by the electroweak precision observables
of the $Z$ boson~\cite{Larios:2004mx}. 
%The $tqH$ coupling contributes to the $Zc\bar{c}$
%vertex  and the $D^0-\bar{D}^0$ mixing  at the loop level. 
The ATLAS and the CMS collaborations have 
set upper limits on the flavor changing neutral currents in the top sector  
through the top pair production, with one top decaying to $Wb$ and the other 
top assumed to decay to $qH$. The leptonic decay mode of the $W$ is considered
and the different Higgs decay channels are analyzed,  
with the Higgs decaying either to two photons~\cite{Aad:2014dya,CMS:2014qxa} 
or to $b\bar{b}$~\cite{Aad:2015pja,CMS:2015qhe}.
Combining the analysis of the different Higgs decay channels,
based at $\sqrt{s}$ = 8 TeV and 
an integrated luminosity of 20.3 (19.7) fb$^{-1}$, the 95\% CL upper limits
obtained by ATLAS (CMS)~\cite{Aad:2015pja,Khachatryan:2016atv} are
Br$(t\rightarrow cH) \leq 4.6 (4.0) \times 10^{-3}$ 
and Br$(t\rightarrow uH) \leq 4.5 (5.5) \times 10^{-3} $.  On the phenomenological 
side the sensitivity of LHC measurements to these non-standard flavor violating couplings in 
the top sector has been explored in great details, considering ($a$) the 
top quark pair production~\cite{AguilarSaavedra:2000aj, Atwood:2013ica, Kobakhidze:2014gqa, Wu:2014dba}, 
($b$) the single top $+$ Higgs production~\cite{AguilarSaavedra:2004wm, Greljo:2014dka}
 and ($c$) single top $+$ W production~\cite{Liu:2016dag}. 

The analysis of the $tqH$ coupling has also been carried out in the context 
of the next generation $e^-e^+$ linear colliders, the International Linear Collider (ILC)
and the Compact Linear Collider (CLIC)~\cite{Behnke:2013xla, Aicheler:2012bya}. 
These planned high energy $e^-e^+$ colliders are expected to perform 
precision measurements of the top-quark and the Higgs boson.
They will be able to scrutinize the couplings
in the top-Higgs sector to the extreme precision, making them suitable for the sensitive tests of physics beyond the SM. 
The baseline machine design for both colliders allows for up to
$\pm 80 \%$ electron polarization, while provisions have been made to allow positron polarization of $\pm 30 \%$ as 
an upgrade option. Both these machines are designed to operate at centre of mass energies of 350, 500 
and 1000 GeV, with the possibility of CLIC to be also adapted for the 3 TeV operation. Several studies have been carried out in the context of zero beam polarization at the ILC~\cite{Han:2001ap, Hesari:2015oya} in an attempt
to constrain the $tqH$ vertex. 
 
The Higgs boson within the SM couples similarly to 
$\bar{q}_Lq_R$ and $\bar{q}_Rq_L$, i.e. $y_{LR}=y_{RL}$. Most of the studies 
in the context of the FCNC in the Higgs sector takes into effect this 
consideration and assumes the similarity between the chiral 
couplings.  In this work we have focussed on the chiral nature of the FCNC 
couplings and have shown how the inequality of chiral couplings leads to distinct behaviour 
in the distributions of final states at linear colliders. 
We work in the context of initial beam polarization for both the electron and the positron, 
using the advantages of their adjustment for enhancing the sensitivities of the measured branching ratios and 
the asymmetries on the FCNC parameters. We also present the results in the case of transverse polarized beams. 

%The helicities of the inital beams 
%being directly coupled to the helicities of the produced tops, we will also use this %effect to study the FCNC couplings. 

It is a well known fact that by a detailed study of the top (antitop)
decay products one can obtain valuable information about the top spin
observables and then use them for the detailed exploration of the top quark
pair production or decay 
dynamics to distinguish among different models of new physics ~(\cite{KamenikMelic} and references therein). In order to maximize the top spin effects it is advisable to 
choose a proper spin quantization axis. 
At the Tevatron, where the top quark pair production was
dominated by the quark-antiquark annihilation a special off-diagonal axis was shown to exist~\cite{Parke:1996pr},
making top spins $100\%$ correlated. On the other hand, at the LHC 
the top quark pair production is dominated by the gluon-gluon fusion and 
there is no such optimal axis for this process\footnote{At 
low $m_{t\bar t}$ the top quark pair production via gluon-gluon fusion is 
dominated by like-helicity gluons. Consequently, spin correlations are 
maximal in the helicity basis~\cite{Mahlon:2010gw}.}. The $t\bar{t}$ 
production through the electron-positron annihilation at the linear 
colliders will be similar to the Tevatron production, therefore
the top quark spins will also be maximally correlated in the off-diagonal basis. 
The $t$, $\bar{t}$ spin effects, can be analyzed in the lepton-lepton 
or lepton+jets final states through a number of angular distributions and 
correlations. The spin information is proportional
to the spin analyzing power of the decay products of the top and will
therefore differ from the SM one in the case of FCNC top-Higgs decay.
We therefore also carry out a detailed study of the FCNC $t\to qH$ decay 
with different spin observables, and in different top-spin polarization basis, 
using both unpolarized and longitudinally polarized beams. 

The outlay of the paper is as follows. We discuss in Sec.~\ref{sec:FCNC}, 
the most general FCNC lagrangian considered for our analysis.
We give a brief review of the effects of initial beam polarizations in the $t\bar{t}$ production at the linear 
collider in Sec.~\ref{sec:beam_pol}. The detailed analysis of the $tqH$
final state is performed in Sec.~\ref{sec:cmframe} and constraints are 
obtained from angular asymmetries. The top spin observables in the context of 
different spin bases are discussed in Sec.~\ref{sec:spintop}. A thorough numerical
study of the process $e^-e^+\rightarrow t\bar{t}\rightarrow qHW^-\bar{b}\rightarrow 
qb\bar{b}l^-\bar{\nu}_l\bar{b}$ including top FCNC coupling is performed in 
Sec.~\ref{sec:numericalstudy}, and finally we conclude in Sec.~\ref{sec:conclusion}.
The analytical form of the different production and decay matrices, along with the expressions for the 
top spin observables used for our analysis are listed in Appendix~\ref{sec:All_appen}
and \ref{sec:appen_spin}.

\section{The flavor changing top quark coupling}\label{sec:FCNC}

We concentrate on the most general FCNC $tqH$ Lagrangian of the form 
\begin{eqnarray}\label{eq:tqhA}
{\cal L}^{tqH} &=& g_{tu} \bar{t}_R u_L H + g_{ut} \bar{u}_R t_L H +  g_{tc} \bar{t}_R c_L H 
+ g_{ct} \bar{c}_R t_L H +  h.c \nonumber \\
&=& \bar{t} (g_{tq} P_L + g_{qt}^\ast P_R ) q H  + \bar{q} (g_{qt} P_L + g_{tq}^\ast P_R ) t H.
\end{eqnarray}   
This Lagrangian gives rise to the tree-level FCNC decays $t \to H q,(q=u,c)$ with the partial decay width given as 
\begin{eqnarray}\label{eq:top_TDW}
\Gamma_{t\rightarrow q H}&=&\frac{1}{32 \pi m_t^3}\sqrt{m_t^2-(m_q-m_H)^2}
\sqrt{m_t^2-(m_q+m_H)^2}\left[(\mid g_{tq}\mid^2 + \mid g_{qt}\mid^2)(m_t^2+m_q^2-m_H^2) \right. \nonumber \\
&& \left. +4 m_t m_q 
\left(g_{tq}^\ast g_{qt} + g_{qt}^\ast g_{tq}\right)\right].
\end{eqnarray}
The SM top-quark decay is dominated by $t\rightarrow b W^+$ and it is given by
\begin{equation}\label{eq:tWb}
 \Gamma_{t\rightarrow b W^+} = \frac{G_F}{8 \sqrt{2} \pi m_t^3}(m_t^2-m_W^2)^2(m_t^2+2 m_W^2) \,.
\end{equation}
We neglect the mass of the emitted quark $m_q$, in our 
analysis. The branching ratios of the top decaying in the presence of these 
flavor violating Yukawa couplings is then given by
\begin{eqnarray}\label{eq_BR}
 {\rm BR}(t\rightarrow q H) &=& \frac{1}{2\sqrt{2}G_F}\frac{(m_t^2-m_H^2)^2}{(m_t^2-m_W^2)^2 (m_t^2+2 m_W^2)}
 (\mid g_{tq}\mid^2 + \mid g_{qt}\mid^2) \alpha_{QCD},
    \end{eqnarray}
where the NLO QCD corrections to the SM decay width \cite{Li:1990qf} and the $t \to c H$ 
decay \cite{Zhang:2013xya} are included in the factor $\alpha_{QCD} = 1+0.97 \alpha_s = 1.10$ \cite{Greljo:2014dka}. 
The total decay width of the top in the presence of these FCNC couplings is then 
\begin{eqnarray}
 \Gamma_t &=& \Gamma_t^{SM} + \Gamma_{t\rightarrow q_H} 
  \approx \Gamma_t^{SM} + 0.155 (\mid g_{tq}\mid^2 + \mid g_{qt}\mid^2) \,.
\end{eqnarray}
We have $\Gamma_{t}^{SM} = \Gamma_{t\rightarrow W^+ b}$ = 1.35 GeV  
for $m_t$ = 173.3 GeV at NLO, while the experimentally observed value of the total 
top-quark  width is, $\Gamma_t = 1.41^{+0.19}_{-0.15}$ GeV~\cite{Agashe:2014kda}.
The additional FCNC decay processes give positive contributions to $\Gamma_t$, proportional to 
$(\mid g_{tq}\mid^2 + \mid g_{qt}\mid^2)$ and from the experimentally observed 
$\Gamma_t$ an upper bound on $\sqrt{\mid g_{tq}\mid^2 + \mid g_{qt}\mid^2}$ 
can be obtained. 
These  flavor changing couplings can also lead to the 
three body decay $h\rightarrow t^\ast (\to  W^+ b)\bar{q}$, where top is
produced off-shell and $q=u,~c$. Then total width of the Higgs gets
modified and the couplings  $g_{tq},~g_{qt}$ can be 
independently constrained from the measurement of the Higgs 
decay width at the LHC~\cite{Atwood:2013ica}.
 
\section{Polarized beams in $t\bar{t}$ production at the $e^- e^+$ linear collider}\label{sec:beam_pol}

The most general formula for the matrix element square $|T_{e^-e^+}|^2$ for arbitrary polarized $e^-e^+$ beams producing a $t\bar{t}$ pair 
is given in Refs.~\cite{Kleiss:1986ct, Hikasa:1985qi}. However for the annihilation process 
with massless electron and  positron, the helicity of the electron has to be opposite to 
that of the positron, and the final formula is reduced to the form,
\begin{eqnarray}
 |T|^2 &=&  \frac{1}{4}\left\lbrace(1-P^L_{e^-})(1+P^L_{e^+}) |T_{e^-_L e^+_R}|^2 + 
 (1+P^L_{e^-})(1-P^L_{e^+}) |T_{e^-_R e^+_L}|^2 \right. \nonumber \\
 && \hspace*{0.5cm}\left. + P^T_{e^-} P^T_{e^+} {\rm Re} \left[  e^{-i(\alpha_- + \alpha_+)} 
 T_{e^-_R e^+_L} T_{e^-_L e^+_R}^{\ast}+ e^{i(\alpha_- + \alpha_+)}T_{e^-_L e^+_R} T_{e^-_R e^+_L}^{\ast}\right ]
 \right\rbrace,
 \label{eq:fin2a}
\end{eqnarray}
where $T_{e^-_{\lambda_1} e^+_{\lambda_2}}$ is the helicity amplitude for the process under
consideration, and $\lambda_1,~\lambda_2$ are the helicities of the electron and the positron,
respectively. $P^L_{e^\mp}$ is the degree of the longitudinal polarization and $P^T_{e^\mp}$ 
is the transversal polarization for the electrons and positrons. 
The $\alpha_{\mp}$ refers to the angle of polarization of the electron and the positron, 
respectively.  The polarizations of the electron and the positron at the linear colliders
are independent and can be arbitrarily changed. The proposed linear colliders (ILC and CLIC)
assume that the following polarizations can be achieved\footnote{It is important to note 
the role of the beam polarization in the $t\bar{t}$ production. For the $-80\%$ of the electron
polarization and $+30\%$ of the positron polarization the initial stated will be dominantly 
polarized as $e^-_L e^+_R$, giving in the SM
a constructive interference of the $\gamma$ and $Z$ amplitudes
for the production of $t_L\bar{t}_R$ pair, and a destructive interference for the 
production of $t_R \bar{t}_L$, which then leads to a large positive forward-backward asymmetry. 
} 
\begin{eqnarray}
P^{L,T}_{e^-} = \pm 80\% \;, P^{L,T}_{e^+} = \pm 30\% \,.
\end{eqnarray}
As it was shown in Ref.~\cite{Hikasa:1985qi}, if one is interested in the $\phi_t$ (azimuthal angle of 
the top quark) dependence of the cross section,  instead of discussing $\phi_t$ dependence directly, 
it is simpler to study $\alpha_{\mp}$ dependence, since the latter is explicit in above. It can be shown that 
\begin{eqnarray}
 | \langle f(\phi_t,...) | T | e^-(\alpha_-) e^+ (\alpha_+) \rangle |^2 = | \langle f(\phi_t = 0,...) | T |
 e^-(\alpha_- - \phi_t) e^+ (\alpha_+ - \phi_t) \rangle |^2 \,,
\end{eqnarray}
from the rotational invariance with respect to the beam direction, i.e. the rotation of the final state 
by $\phi_t$ is equivalent to the rotation of the initial state by $-\phi_t$. %Additional simplification can be, for $\alpha_{\pm} \rightarrow \alpha_{\pm} - \phi$ above, a choice 
%that $\alpha_- =0$. 
With this assumption Eq.~(\ref{eq:fin2a}) becomes 
\begin{eqnarray}
 |T|^2 &=&  \frac{1}{4}\left\lbrace(1-P^L_{e^-})(1+P^L_{e^+}) |T_{e^-_L e^+_R}|^2 + 
 (1+P^L_{e^-})(1-P^L_{e^+}) |T_{e^-_R e^+_L}|^2 \right. \nonumber \\
 && \hspace*{0.5cm}\left. -2 P^T_{e^-} P^T_{e^+} 
\rm{Re}~e^{i(\eta-2\phi_t)} T^{\ast}_{e^-_R e^+_L} T_{e^-_L e^+_R}
 \right\rbrace,
 \label{eq:fin2}
\end{eqnarray}
where $\eta = \alpha_-+\alpha_+$.
The effects of various beam polarizations in above will be discussed in the following.  
 
\section{Analysis of the $tqH$ final state at the $e^- e^+$ linear collider}\label{sec:cmframe}

We study the $t\bar{t}$ production in the context of
the $e^-e^+$ linear collider, where one of the top decays to $Wb$, and the 
other decays to $q(u,c)H$ and the leptonic decay mode of 
the $W$ boson is considered: %The process is shown in Fig.~\ref{fig:processtt}, 
\begin{eqnarray}
&&  e^-(p_1) + e^+(p_2)\rightarrow t(q_1)+\bar{t}(q_2), \nonumber \\
&&  \hspace*{3cm} t(q_1) \rightarrow q (p_{q})+ H,  \qquad 
\bar{t}(q_2) \rightarrow \bar{b} (p_b)+ l^+(p_l)+ \nu(p_\nu).
\end{eqnarray}
We first consider the leading order spin dependent differential cross-section of the top pair production
in a generic basis. The total phase space is split into the product of the differential cross-section for 
the $t\bar{t}$ production, the three-particle decay of the antitop quark  and the two-particle decay 
of the top quark, with the Higgs decaying to $b\bar{b}$. We first do the analysis considering the decay of $t$ to $qH$ and the inclusive decay of $\bar{t}$. In an attempt to make a comparative study, we also consider the $t\bar{t}$ production, with the SM decay of top to $W^+b$,
and the inclusive decay of $\bar{t}$. This SM process will be a background for the $tqH$ final state, with the $H$ 
and the $W$ decaying hadronically. 
Since the analysis is being similar for both, the considered signal and the SM 
background, we only discuss the calculation of the signal in details.
The differential cross section in the centre of mass frame becomes
\begin{eqnarray}\label{df_cs1}
d\sigma &=& \frac{1}{2s} \int \frac{ds_1}{2\pi} \frac{1}{((s_1-m_t^2)^2+\Gamma_t^2m_t^2)}
 \times \mid \bar{\mathcal{M}}^2 \mid \nonumber \\
&\times&  (2 \pi)^4 \delta^4 (q_1+q_2-p_1-p_2) \frac{d^3q_1}{(2\pi^3) 2E_1} \frac{d^3q_2}{(2\pi^3) 2E_2}~~~~~
[\rm{production~of}~ t\bar{t}] \nonumber \\
&\times&  (2\pi)^4 \delta^4(p_q+p_H-q_1) \frac{d^3p_q}{(2\pi^3) 2E_q} \frac{d^3p_H}{(2\pi^3) 2E_H} 
~~~~~[\rm{decay~ of}~ t]\,,
\end{eqnarray}
where $\sqrt{s}$ is the centre of mass energy and $s_1 = (p_q+p_H)^2$. The energies of the produced top and the antitop are given by 
$E_1,~E_2$, whereas the energies of the decay products are denoted by $E_q$ and $E_H$. 
For these decays, in the center of mass frame and in the narrow width approximation, we can express the elements of the phase space in (\ref{df_cs1}) as
\begin{eqnarray}
&&\int \frac{ds_1}{2\pi} \frac{1}{((s_1-m_t^2)^2+\Gamma_t^2m_t^2)}
 = \int \frac{ds_1}{2\pi} \frac{\pi}{m_t \Gamma_t} \delta(s_1-m_t^2)
 =\frac{1}{2 m_t \Gamma_t} \,,\label{t_prop} \\ 
&&\int \frac{1}{2s} (2 \pi)^4 \delta^4 (q_1+q_2-p_1-p_2) \frac{d^3q_1}{(2\pi^3) 2E_1} \frac{d^3q_2}{(2\pi^3) 2E_2}
 = \frac{3\beta}{64 \pi^2 s} d\cos\theta_t d\phi_t  \,, \label{prod_tt} \\
&&\int (2\pi)^4 \delta^4(p_q+p_H-q_1) \frac{d^3p_q}{(2\pi^3) 2E_q} \frac{d^3p_H}{(2\pi^3) 2E_H}=\frac{1}{2(2\pi)^2}
\int d\Omega_q \frac{\mid p_q \mid^2}{(m_t^2-m_H^2)} \,. 
\label{t_decay}
\end{eqnarray}
The total matrix element squared $\mid \bar{\mathcal{M}}^2 \mid $ in Eq.~(\ref{df_cs1}), 
is defined as
\begin{eqnarray}
\mid \bar{\mathcal{M}}^2 \mid&=& \sum_{L,R}\sum_{(\lambda_t\lambda_t'=\pm)}\rho^{P(t\bar{t})}_{LR,\lambda_t\lambda_t'}\rho^{D(t)}_{\lambda_t\lambda_t'} = \sum_{L,R}\sum_{(\lambda_t\lambda_t'=\pm)}
 \mathcal{M}^{L,R}_{\lambda_t}\mathcal{M}^{*L,R}_{\lambda_t'}
 \rho^{D(t)}_{\lambda_t\lambda_t'},
\end{eqnarray}
where $ \mathcal{M}^{L,R}_{\lambda_t}$ is the production helicity amplitude of the top with a given helicity 
$\lambda_t$. The helicities of the antitop are summed over. The production helicity amplitudes are listed
in Eqs.(\ref{va_hel}) of Appendix \ref{sec:Appendix1}. 
The decay matrix of the top quark is defined as $\rho^{D(t)}_{\lambda_t\lambda_t'}=\mathcal{M}(\lambda_t)
\mathcal{M}^*(\lambda_t')$ and for $t \to q H$ the explicit expressions in the rest frame of the top,  as well as in the centre of mass frame are given in Appendix \ref{sec:appen_decaytcH}. For the top decaying to $W^+b$ the spin density matrix 
$\rho^{D(t)}_{\lambda_t\lambda_t'}$, is given in Appendix~\ref{sec:appen_decaytWb}, 
for both the top rest frame and the centre of mass frame. 

\begin{figure}[htb]
\centering
\includegraphics[width=10cm, height=6.5cm]{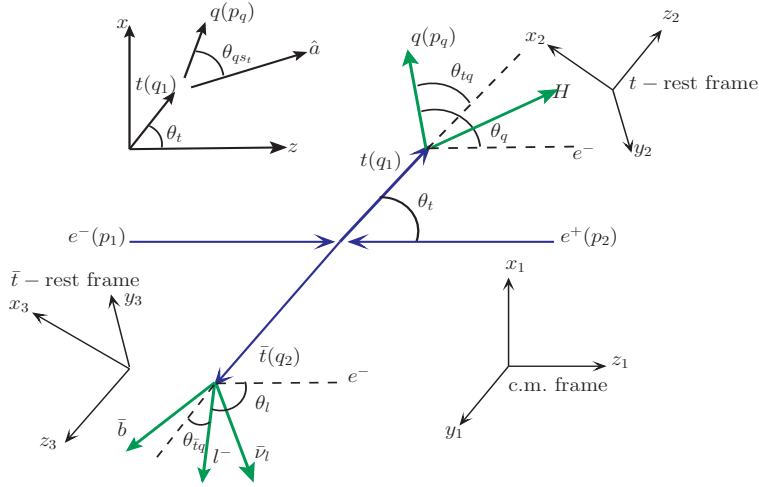}
\caption{The coordinate system in the colliding $e^-e^+$ centre of mass 
frame. The $y$-axis is chosen along the $p_1 (e^-) \times q_1 (t)$ direction and is pointing towards
the observer. The coordinate systems in the $t$ and $\bar{t}$ rest frames are obtained from it by rotation 
along the $x$ axis and then boost along the $y$ axis.}
\label{fig:eett_2nd_frame}
\end{figure}

We have performed our calculations, in the frame where the electron beam direction  
is in the positive $z$ direction, with the top emitted at a polar angle 
$\theta_t$ and the quark emitted in the top decay makes a polar $\theta_q$ angle with
the electron beam, as 
shown in Fig.~\ref{fig:eett_2nd_frame}. 
The four-vector in the rest frame of the top are related to the c.m. frame by the following 
boost and the rotation matrices (the boost matrix is along the $z$ direction, whereas the 
rotation matrix is applied along the $y$ axis):
\begin{eqnarray}
 q_1&=&\begin{pmatrix}
          1&0&0&0 \\ 0&\cos\theta_t&0&\sin\theta_t \\ 0&0&1&0 \\ 0&-\sin\theta_t &0&\cos\theta_t
         \end{pmatrix} \begin{pmatrix} 
                \gamma &0&0&\gamma\beta \\ 0&1&0&0 \\0&0&1&0 \\ \gamma\beta &0&0&\gamma            
            \end{pmatrix} q_1^{top}   \,,   
\end{eqnarray}
where $q_1^{top}$ is defined in the rest frame of the top. 
The momentum four-vectors in the c.m frame are given by
\begin{eqnarray}
p_1 &=& \frac{\sqrt{s}}{2}(1,0,0,1),~~~~p_2 = \frac{\sqrt{s}}{2}(1,0,0,-1) \nonumber \\
q_1&=& \frac{\sqrt{s}}{2}(1,\beta \sin\theta_t,0,\beta\cos\theta_t),~~~~
q_2= \frac{\sqrt{s}}{2}(1,-\beta \sin\theta_t,0,-\beta\cos\theta_t)  \nonumber \\
p_{q}&=&(E_{q}, E_{q}\sin\theta_{q} \cos\phi_{q}, E_{q}\sin\theta_{q}\sin\phi_{q}, E_{q}\cos\theta_{q})
\end{eqnarray}
The momentum of the emitted light quark $\mid p_q \mid$ is equal to its energy 
$E_{q}$
and in the c.m frame the following relations are obtained: 
\begin{eqnarray} \label{eq:cosTHtq}
\mid p_q \mid &=& E_{q} = \frac{(m_t^2-m_H^2)}{\sqrt{s} (1-\beta \cos\theta_{tq})}, \nonumber \\
\cos \theta_{tq} &=& \cos\theta_t \cos \theta_q +\sin\theta_t \sin \theta_q \cos \phi_q \,. 
\end{eqnarray}
where $\cos \theta_{tq}$ is the angle between the top and the emitted light quark in the c.m. frame.

Combining the production and the density matrices in the narrow width approximation for $t$, we get the 
polar distribution of the emitted quark $q$, in the presence of the beam polarization
after integrating over $\phi_q,\theta_t$, to be 
\begin{eqnarray}
 \frac{d\sigma}{ds~d\cos\theta_q~d\phi_t} &=& |T|^2
%\frac{1}{4}\left((1-P^L_{e^-})(1+P^L_{e^+}) |T_{e^-_L e^+_R}|^2 + 
% (1+P^L_{e^-})(1-P^L_{e^+}) |T_{e^-_R e^+_L}|^2\right) \nonumber \\
% &&-\frac{1}{2} P^T_{e^-} P^T_{e^+} \rm{Re}~e^{i(\eta-2\phi_t)} T^{\ast}_{e^-_R e^+_L} T_{e^-_L e^+_R} \,,
\end{eqnarray}
where $|T|^2$ is of the form given in Eq.~(\ref{eq:fin2}). We compute $|T_{e^-_L e^+_R}|^2$, $|T_{e^-_R e^+_L}|^2$
for the considered process, and present them in the most general form:
\begin{eqnarray}
 |T_{e^{\mp}_L e^{\pm}_R}|^2 &=& (|g_{tq}|^2+|g_{qt}|^2) \left(a_0 + a_1 \cos\theta_q  + a_2 \cos^2\theta_q \right) \nonumber \\ 
&& \pm(|g_{tq}|^2-|g_{qt}|^2) \left(b_0+ b_1 \cos\theta_q + b_2 \cos^2\theta_q \right) \,.
\end{eqnarray}
The coefficients $a_{i},b_{i}$ 
%are coefficients of $\cos^{i}\theta_q$, $i=0,1,2$ and 
can be deduced from the following expressions: 
\begin{align}\label{eq:Long_mp}
 |T_{e^-_L e^+_R}|^2 &= (m_t^2-m_H^2) \frac{\pi s}{\beta} \left \lbrace \frac{|g_{tq}|^2+|g_{qt}|^2}{\beta^2-1}
 \left[-4 A_L B_L \cos\theta_q \left ( \beta +(\beta^2-1)\tanh^{-1}\beta\right)  \right. \right. \nonumber \\
 & \left. \left.+ (A_L^2+B_L^2) \cos^2\theta_q \left (\beta(2\beta^2-3)-3(\beta^2-1)\tanh^{-1} \beta\right )
  \right. \right. \nonumber \\
 & \left. \left. +\left(-\beta +(\beta^2-1)\tanh^{-1}\beta\right)(A_L^2+B_L^2) -2\beta(\beta^2-1)B_L^2 \right]
 \right.  \nonumber \\
 & \left. + 2(|g_{tq}|^2-|g_{qt}|^2)\left[\cos\theta_q \left((A_L^2+B_L^2)\tanh^{-1} \beta -\beta B_L^2 \right)
 \right. \right. \nonumber \\ 
& \left. \left. 
+A_L B_L 
 \left(1-3 \cos^2\theta_q\right) (\beta-\tanh^{-1} \beta) \right] \right\rbrace, \\
 |T_{e^-_R e^+_L}|^2 &= (m_t^2-m_H^2) \frac{\pi s}{\beta} \left \lbrace \frac{|g_{tq}|^2+|g_{qt}|^2}{\beta^2-1}
 \left[-4 A_R B_R \cos\theta_q \left ( \beta +(\beta^2-1)\tanh^{-1}\beta\right )  \right. \right. \nonumber \\
 & \left. \left.+ (A_R^2+B_R^2)\cos^2\theta_q \left ( 
 \beta(2\beta^2-3)-3(\beta^2-1)\tanh^{-1} \beta\right )
  \right. \right. \nonumber \\
 & \left. \left. +\left(-\beta +(\beta^2-1)\tanh^{-1}\beta\right)(A_R^2+B_R^2) -2\beta(\beta^2-1)B_R^2 \right]
 \right.  \nonumber \\
 & \left. - 2(|g_{tq}|^2-|g_{qt}|^2)\left[\cos\theta_q \left((A_R^2+B_R^2)\tanh^{-1} \beta -\beta B_R^2 \right)
 \right. \right. \nonumber \\ 
& \left. \left. +A_R B_R 
 \left(1-3 \cos^2\theta_q\right) (\beta-\tanh^{-1} \beta) \right] \right\rbrace,\label{eq:Long_pm}
\end{align}
and 
%$T^{\ast}_{e^-_R e^+_L} T_{e^-_L e^+_R}$ is given by
\begin{align}\label{eq:Trans}
 T^{\ast}_{e^-_R e^+_L} T_{e^-_L e^+_R} &=  \frac{\pi s}{\beta}
(m_t^2-m_H^2) (\beta-\tanh^{-1} \beta) (3\cos^2\theta_q-1) \cos (\eta-2\phi_t) \nonumber \\
&\left\lbrace  (|g_{tq}|^2+|g_{qt}|^2)(A_L A_R-B_L B_R)
+(|g_{tq}|^2-|g_{qt}|^2)(A_L B_R -A_R B_L)\right\rbrace,
 \end{align}
where $A_{L,R}$ and $B_{L,R}$ are combinations of the standard SM $\gamma$ 
and $Z$ couplings with the top and the leptons in the $t\bar{t}$ production given 
in the Eq.~(\ref{eq:albl}). The Yukawa chiral couplings, as seen from 
Eqs.~(\ref{eq:Long_mp}),~(\ref{eq:Long_pm}) are both proportional 
to the polar angle of the emitted light quark, $\cos\theta_q, \cos^2\theta_q$, but 
have different dependencies. The coefficients of the coupling $|g_{tq}|^2$, which 
measures the coupling strength of $t_L$ with $q_R$ and the Higgs, are summed 
in Eq.~(\ref{eq:Long_mp}), whereas the coefficients of the other chiral coupling
$|g_{qt}|^2$ do not add up, but cancel each other partially. This is the case 
when the electron beam is left polarized and the positron is right polarized. 
This behaviour of $|g_{tq}|^2$ and $|g_{qt}|^2$ is reversed 
with the right polarized electrons and the left polarized positrons, as can be noticed from Eq.~(\ref{eq:Long_pm}), where the coefficients of $|g_{qt}|^2$ add up. Therefore, 
it will be possible to control the influence of particular chiral couplings with a
suitable choice of beam polarization. The case of transverse polarization is also 
considered, although both $|g_{tq}|^2,~|g_{qt}|^2$ involve same angular dependencies in 
Eq.~(\ref{eq:Trans}) and therefore cannot be used for the analysis of the chirality of the FCNC couplings. 
%functions of 
%$\theta_q$ and $\phi_t$ such as $\cos(\eta-2\phi_t)$,
%and $\cos(\eta-2\phi_t)\cos^2\theta_q$. 
It is clear from Eqs.~(\ref{eq:Long_mp}),~(\ref{eq:Long_pm}),~(\ref{eq:Trans}), that $|g_{tq}|^2$ and $|g_{qt}|^2$ cannot be isolated separately, but 
their effects can be individually controlled with suitable choice of beam 
polarization. We next study different distributions in the presence of the 
chiral FCNC couplings and accordingly construct asymmetries to set limits on them.

\subsection{Constraints on the chiral FCNC couplings by angular asymmetries}\label{sec:cmf_asymmetries}

Next, we perform a detailed analysis of the signal FCNC process considered, along with the standard SM background ($t\bar{t}, t\rightarrow Wb$, W decaying hadronically) and construct different asymmetries for obtaining limits on the couplings. 

The total cross section for both the signal and
the background, in case of the longitudinal beam polarization is 
\begin{eqnarray}
\sigma_{Signal} &=& \frac{(m_t^2-m_H^2)^2}{4 s \Gamma_t m_t}\frac{1}{1-\beta^2}(|g_{tq}|^2+|g_{qt}|^2)\left((1-P^L_{e^-})(1+P^L_{e^+})
\left(s \beta^2  B_L^2 + (2 m_t^2+s) A_L^2\right)  \right. \nonumber \\
&& \left. + (1+P^L_{e^-})(1-P^L_{e^+}) \left(s \beta^2 B_R^2  + (2 m_t^2+s) A_R^2  \right) \right)\,, \\
\sigma_{Bkg} &=&\frac{g^2 m_t}{2 s^2 \Gamma_t m_W^2}\frac{1}{(1-\beta^2)^2} (m_t^2-m_W^2)^2(m_t^2+2m_W^2)
\left((1-P^L_{e^-})(1+P^L_{e^+})
\left(s \beta^2  B_L^2 \right. \right. \nonumber \\ 
&& \left. \left. + (2 m_t^2+s) A_L^2\right)  
 + (1+P^L_{e^-})(1-P^L_{e^+}) \left(s \beta^2 B_R^2  + (2 m_t^2+s) A_R^2  \right) \right)\,, 
\end{eqnarray}
where again $A_{L,R}$ and $B_{L,R}$ are combinations of the SM $\gamma$ and $Z$ couplings with
the quarks in the $t\bar{t}$ production given in Appendix~\ref{sec:All_appen}.

We have performed our analysis considering $\sqrt{|g_{tq}|^2+|g_{qt}|^2}$ = 0.16,
in accordance with the latest LHC bounds~\cite{Greljo:2014dka}.
The background i.e. the SM $\bar{t}Wb$ contribution is scaled down, to be  
compared with the signal. We are currently not applying any cuts on the 
final state, but a detailed analysis using all the experimental cuts will be
performed in Sec.~\ref{sec:numericalstudy}.

\begin{figure}[htb]
  \begin{subfigure}{0.45\linewidth}
  \centering
    \includegraphics[width=6.5cm, height=5cm]{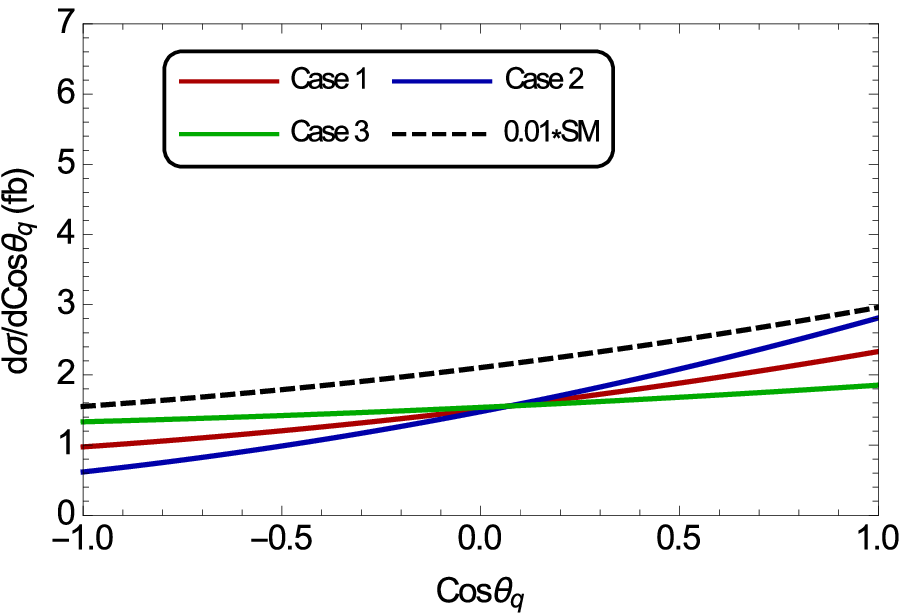}
\caption{}
\label{fig:upa}
  \end{subfigure}
   \hspace{1.0cm}
  \begin{subfigure}{0.45\linewidth}
  \centering
    \includegraphics[width=6.5cm, height=5cm]{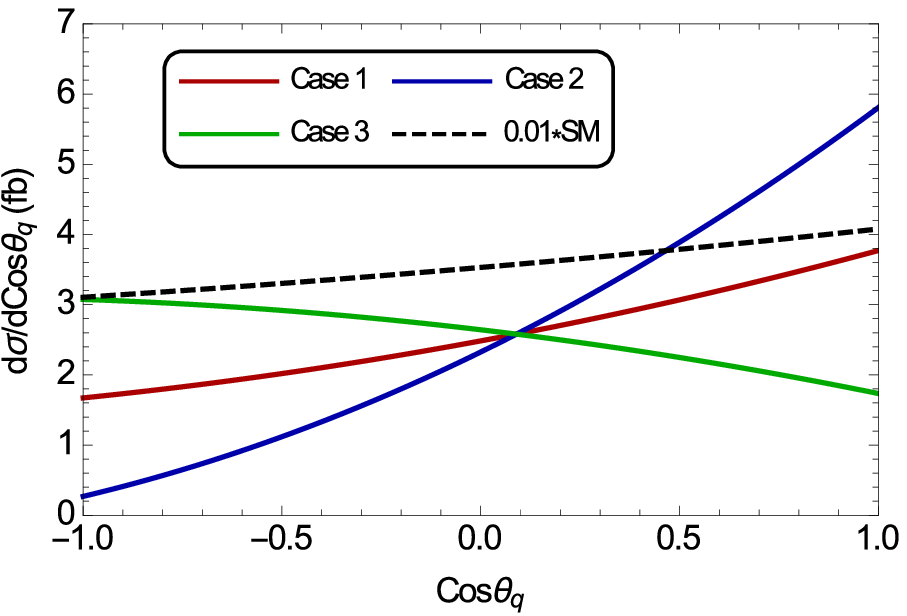}
\caption{}
\label{fig:plr}
  \end{subfigure}\\[1ex]
  \begin{subfigure}{\linewidth}
    \hspace{3.5cm}
  \includegraphics[width=6.5cm, height=5cm]{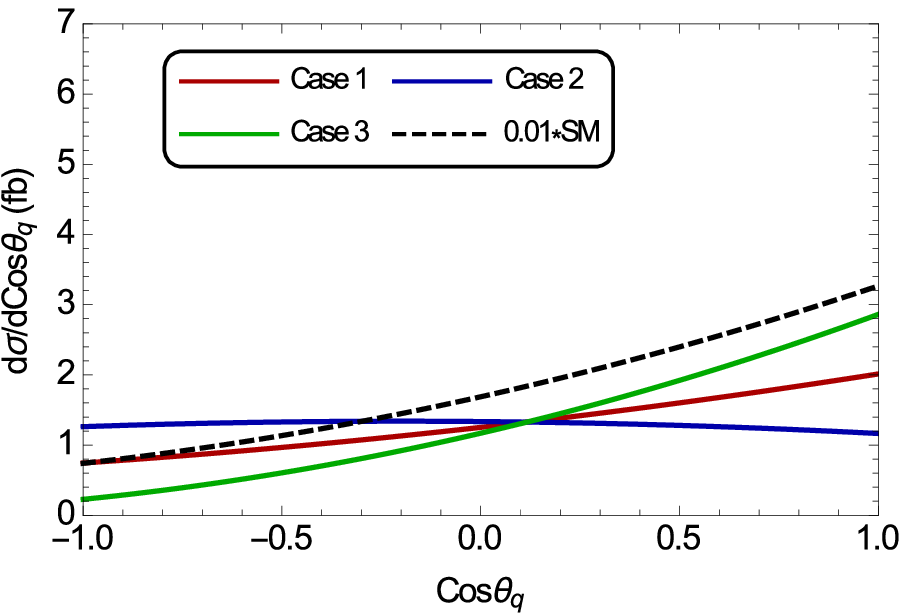}
\caption{}
\label{fig:prl}
  \end{subfigure}
   \caption{The polar angle distribution of the quark at $\sqrt{s}= 500$ GeV,
for ($a$) $P^L_{e^-} = P^L_{e^+} = 0$, ($b$) $P^L_{e^-} = -0.8,~P^L_{e^+} = 0.3$ and 
($c$) $P^L_{e^-} = 0.8,~P^L_{e^+} = -0.3$. The different Cases
are discussed in the text.}
\label{fig:dist_pol}
\end{figure}

The polar angle distribution of the 
emitted quark is plotted in Fig.~\ref{fig:dist_pol} for both, the signal and 
the background, for 
($a$) $P^L_{e^-} = P^L_{e^+} = 0$, ($b$) $P^L_{e^-} = -0.8,~P^L_{e^+} = 0.3$ 
and $c$) $P^L_{e^-} = 0.8,~P^L_{e^+} = -0.3$. The polar 
angle distribution will be  sensitive to the chirality of the Yukawa couplings and 
therefore we present our results for three different cases:
\begin{itemize}
 \item Case 1 : $\sqrt{|g_{tq}|^2+|g_{qt}|^2}$ = 0.16\,,
 \item Case 2 : $\sqrt{|g_{tq}|^2+|g_{qt}|^2}$ = 0.16, with $|g_{qt}|^2$ = 0 \,,
 \item Case 3 : $\sqrt{|g_{tq}|^2+|g_{qt}|^2}$ = 0.16, with $|g_{tq}|^2$ = 0 \,.
\end{itemize}
It can be clearly seen from Fig:~\ref{fig:dist_pol}, that $|g_{tq}|^2$ 
and $|g_{qt}|^2$ are sensitive to the beam polarization. The different Cases
behave similar in the unpolarized case, Fig.~\ref{fig:upa}.
Case 2 is most prominent when the electron beam is left polarized and 
the positron is right polarized, Fig.~\ref{fig:plr}, whereas Case 3 is distinct for 
the scenario with right polarized electrons and 
left polarized positrons, Fig.~\ref{fig:prl}. Therefore the manifestation of the 
dominance of one of the coupling, if present, will be prominent using the 
suitable initial beam polarization.  

\begin{figure}[htb]
  \begin{subfigure}{0.45\linewidth}
  \centering
    \includegraphics[width=6.5cm, height=5cm]{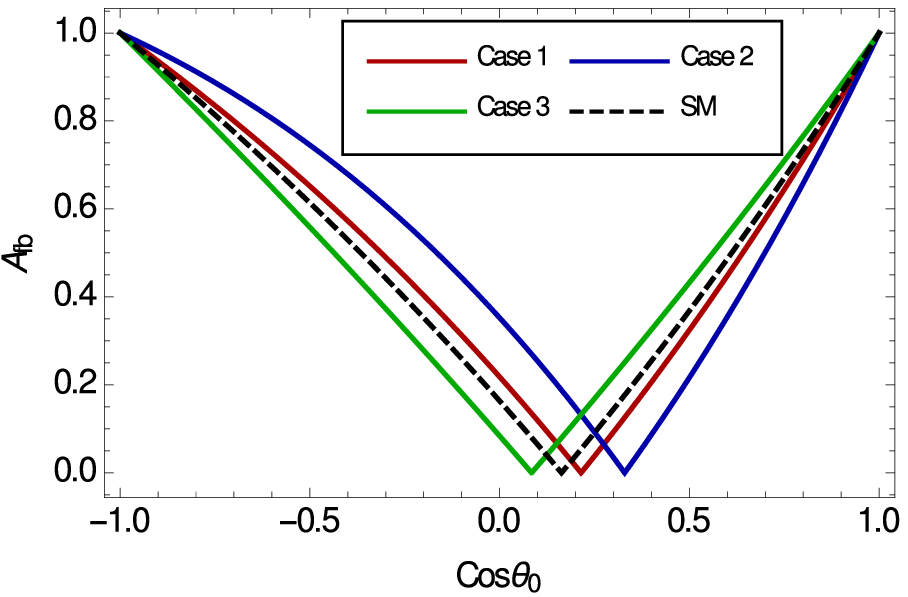}
\caption{}
\label{fig:fb_up}
  \end{subfigure}
   \hspace{1.0cm}
  \begin{subfigure}{0.45\linewidth}
  \centering
    \includegraphics[width=6.5cm, height=5cm]{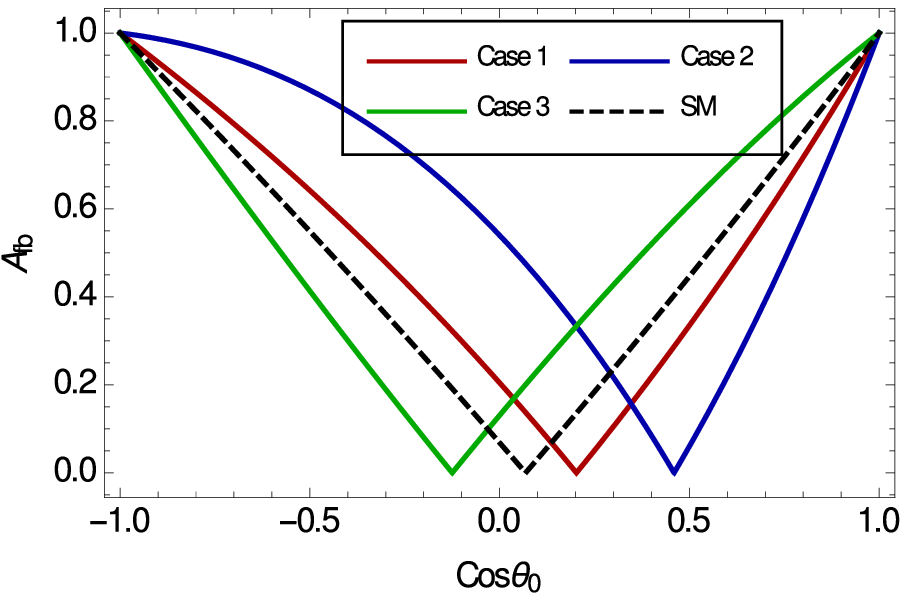}
\caption{}
\label{fig:fb_lr}
  \end{subfigure}\\[1ex]
  \begin{subfigure}{\linewidth}
    \hspace{3.5cm}
  \includegraphics[width=6.5cm, height=5cm]{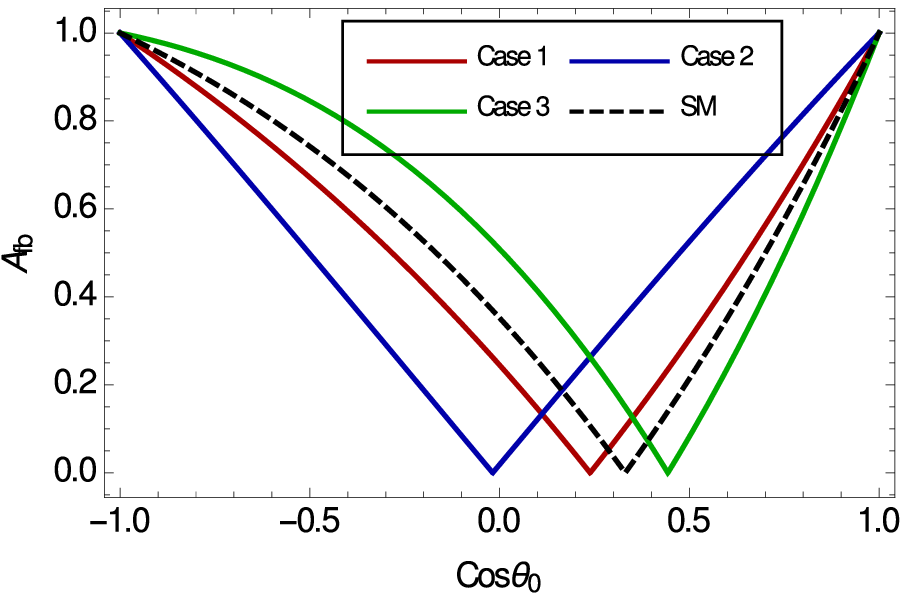}
\caption{}
\label{fig:fb_rl}
  \end{subfigure}
\caption{The forward backward asymmetry as a function of the cut-off angle 
$\cos\theta_0$ Eq.~(\ref{eq:fb}) at $\sqrt{s}= 500$ GeV, for ($a$) $P^L_{e^-} = P^L_{e^+} = 0$,
($b$)$P^L_{e^-} = -0.8,~P^L_{e^+} = 0.3$ and ($c$) $P^L_{e^-} = 0.8,~P^L_{e^+} = -0.3$. The different Cases are discussed in the text.}
\label{fig:FB}
\end{figure}

Using the above fact that the couplings are sensitive to the polar 
angle distributions of the quark, we next consider 
different asymmetries to give simultaneous limits to both of the 
couplings. The $|g_{tq}|^2$ and $|g_{qt}|^2$ terms are accompanied by $\cos\theta_q$,~ $\cos(\eta-2\phi)$ and $\cos(\eta-2\phi) \cos^2\theta_q$ angular 
dependence. The asymmetries which will isolate these terms are the forward-backward asymmetry 
and the azimuthal asymmetry defined as 
\begin{eqnarray}
A_{fb}(\cos\theta_0)&=& \frac{1}{d\sigma/ds}
\left(\int^1_{\cos\theta_0} d\cos\theta_q -  \int^{\cos\theta_0}_{-1} d\cos\theta_q \right)
\frac{d\sigma}{ds~d\cos\theta_q} \,,\label{eq:fb}\\
A_{\phi}(\cos\theta_0) &=& \frac{1}{d\sigma/ds}\left(\int^{\cos\theta_0}_{-\cos\theta_0} d\cos\theta_q
\int_0^{2\pi} d \phi_t~sgn(\cos(\eta - 2\phi_t))\right) \frac{d\sigma}{ds~d\Omega} \,,
\label{eq:trans_asym} 
\end{eqnarray}
where $\theta_0$ is the experimental polar-angle cut \cite{Grzadkowski:2000nx,Rindani:2003av}
and $\Omega = d\cos\theta_q~d\phi_t$. 
The forward-backward asymmetry will isolate the terms proportional to $\cos\theta_q$ in Eqs.(\ref{eq:Long_mp})
and (\ref{eq:Long_pm}). We plot in Fig.~\ref{fig:FB}, the forward backward asymmetry as a function of 
the cut-off angle $\cos\theta_0$. The dip in the plot is where the value of 
$A_{fb}(\cos\theta_0)$ is zero. 
In the presence of $|g_{tq}|^2~(|g_{qt}|^2 = 0)$, i.e Case 2, with left polarized electrons 
and right polarized positrons, the quarks are emitted in the forward direction with the dip 
of $A_{fb}$ to be greater than zero, Fig.~\ref{fig:fb_lr},  
whereas the other Cases almost follow the SM distribution.  
Similarly, with the opposite choice of beam polarization, the 
$|g_{qt}|^2~(|g_{tq}|^2 = 0)$ coupling leads to the quarks being emitted in the forward direction, 
resulting in the dip of $A_{fb}$ to be greater than zero for Case 3 in Fig.~\ref{fig:fb_rl}.  

\begin{figure}[htb]
\centering
\includegraphics[width=6.9cm, height=4.9cm]{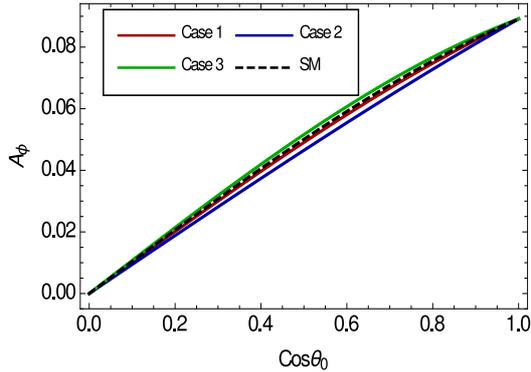}
\caption{The azimuthal asymmetry $A_{\phi}(\theta_0)$ as a function of
$\cos\theta_0$  Eq.~(\ref{eq:trans_asym}) at $\sqrt{s}= 500$ GeV, for the transversal polarizations 
$P^T_{e^-} =0.8$ and $P^T_{e^+} =$ 0.6. The different Cases are discussed in
the text.}
\label{fig:azimuthal}
\end{figure}

Next, we plot the azimuthal asymmetry $A_{\phi}(\cos\theta_0)$ as a function 
of $\cos\theta_0$ in Fig.~\ref{fig:azimuthal}. The terms proportional 
to $\cos(\eta-2\phi_t)$ in Eq.~(\ref{eq:Trans}) survive. We have considered
$\eta$ = 0 for our analysis and $P^T_{e^-} = 0.8$ and $P^T_{e^+} =$ 0.3. The 
distribution is similar for the signal and the background, therefore this 
will not be an useful observable\footnote{ 
However once the FCNC coupling is discovered, this asymmetry can
be used as an additional observable to give limits to the couplings.}.

We compute the limits on the FCNC couplings from the measurement of the forward-backward asymmetry, 
of $e^-e^+ \rightarrow t\bar{t}, t\rightarrow b W^+$ in the SM.
The statistical fluctuation in the asymmetry ($A$), for a given luminosity $\mathcal{L}$ 
and fractional systematic error $\epsilon$, is given as
\begin{eqnarray}
\Delta A^2 &=&\frac{1-A^2}{\sigma \mathcal{L}}+\frac{\epsilon^2}{2}(1-A^2)^2,
\end{eqnarray}
where $\sigma$ and $A$ are the values of the cross section and the asymmetry. The value
of $\epsilon$ is set to zero for our analysis. 
We define the statistical significance of an asymmetry prediction for the new physics,
$A_{FCNC}$, as the number of standard deviations that it lies away from the SM result $A_{SM}$, 
\begin{equation}
 s=\frac{|A_{FCNC}-A_{SM}|}{\Delta A_{SM}}\,,
\end{equation}
where $A_{FCNC}$ is the asymmetry calculated for the process $e^-e^+ \rightarrow t(\to c H) \bar{t}$. 
\begin{figure}[h]
  \begin{subfigure}{0.45\linewidth}
  \centering
    \includegraphics[width=7cm, height=5cm]{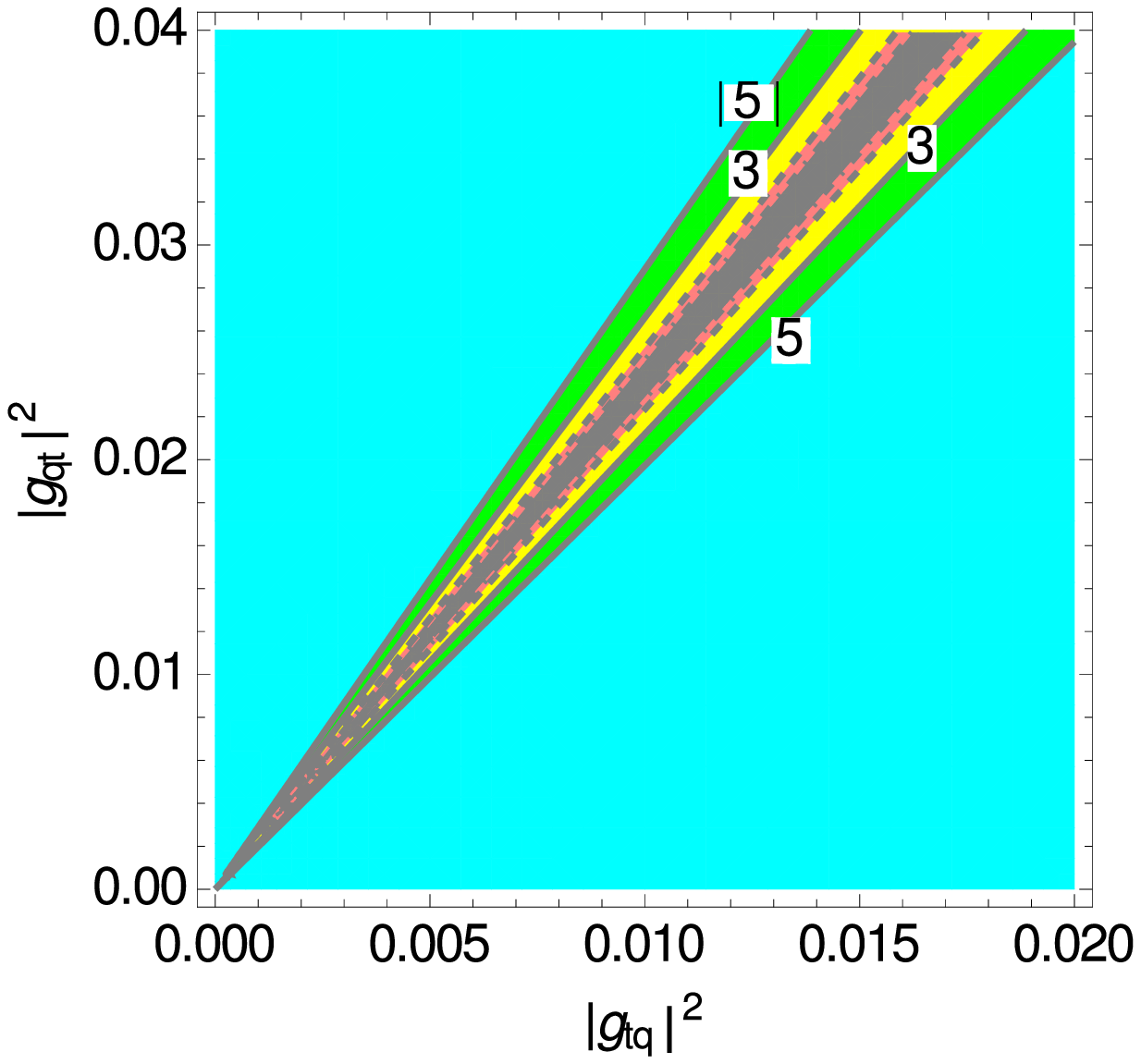}
\caption{}
\label{fig:contour1}
  \end{subfigure}
   \hspace{1.0cm}
  \begin{subfigure}{0.45\linewidth}
  \centering
    \includegraphics[width=7cm, height=5cm]{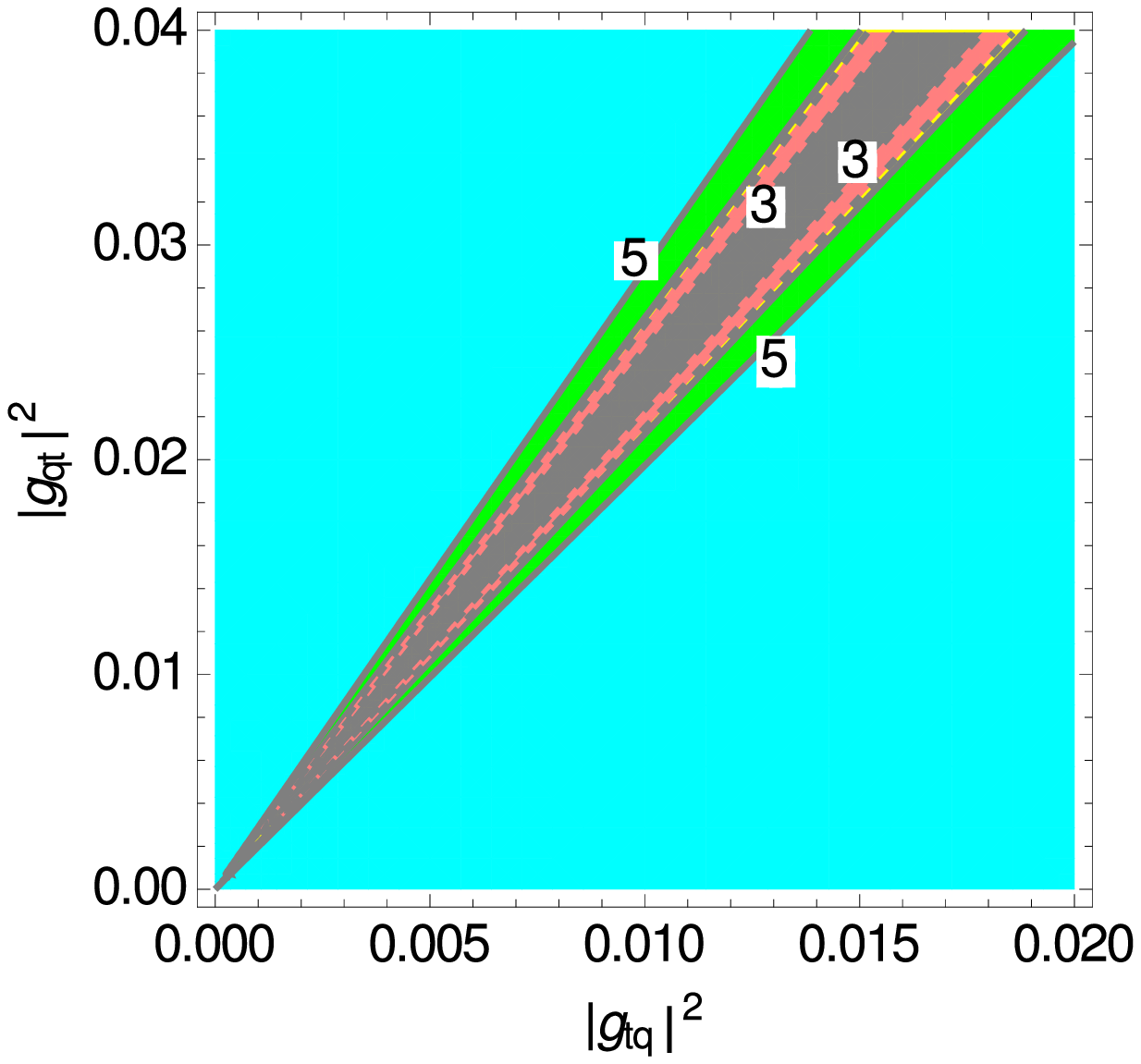}
\caption{}
\label{fig:contour2}
  \end{subfigure}\\[1ex]
\caption{Contour plots of 3$\sigma$ and 5$\sigma$ statistical significance 
in the $|g_{tq}|^2-|g_{qt}|^2|$ region from $A_{fb}$ for $\theta_0$ = 0, Eq.~(\ref{eq:fb})
at $\sqrt{s}$ = 500 GeV and ${\cal L}$ = 500 fb$^{-1}$. 
The solid lines are for the unpolarized case, the dashed lines are for a beam polarization of
($a$) $P^L_{e^-}$ = -0.8, $P^L_{e^+}$ = 0.3 ($b$)  $P^L_{e^-}$ = 0.8, 
$P^L_{e^+}$ = -0.3. Region in blue will be probed at 5$\sigma$ and the green+blue area will be explored at 3$\sigma$
with unpolarized beams. The inclusion of the beam polarization probes yellow+green+blue area 
at 5$\sigma$ and pink+yellow+green+blue at 3$\sigma$. The region which can not be probed
by ILC with this choice of beam polarization is shown in grey.}
\label{fig:contourall}
\end{figure}
We show in Fig.~\ref{fig:contourall} the $|g_{tq}|^2-|g_{qt}|^2$ region, which can be probed 
at a statistical significance of 3$\sigma$ and 5$\sigma$, with both unpolarized and 
polarized beams. The outside area surrounding solid lines can be probed with unpolarized
beams and the outside area surrounding dashed lines can be probed with a beam
polarization of $P_{e^-}^L = -0.8,~P_{e^+}^L = 0.3$ (Fig.~\ref{fig:contour1}),
$P_{e^-}^L = 0.8,~P_{e^+}^L = -0.3$ (Fig.~\ref{fig:contour2}). Obviously, the inclusion of the beam 
polarization can probe a greater region of the $|g_{tq}|^2-|g_{qt}|^2$
parameter space. The $\cos\theta_q$ terms 
%$2(|g_{tq}|^2-|g_{qt}|^2|) (\tanh^{-1}\beta(A_L^2+B_L^2)
%-\beta B_L^2))$ 
in Eqs.~(\ref{eq:Long_mp}-\ref{eq:Long_pm}) 
%and 
%$-2(|g_{tq}|^2-|g_{qt}|^2|) (\tanh^{-1}\beta(A_R^2+B_R^2)-\beta B_R^2))$ 
%in Eq.~(\ref{eq:Long_pm}), 
cancel each other in case of unpolarized beams.
The region in grey is the one, which cannot be explored by ILC with this choice of
the beam polarization.

Now we turn to the discussion of  different top spin observables which can be used to study the FCNC couplings. 

\section{Top spin observables at the ILC}\label{sec:spintop}

We investigate in this section the top spin polarization in the context of the linear collider, 
as the spin information of the decaying top is not diluted by hadronization. In an attempt
to understand the top spin correlations, we work in the zero momentum frame ($t\bar{t}$-ZMF)
\cite{Bernreuther:2004jv} of the $t\bar{t}$ quarks, which is 
\begin{eqnarray}
 (q_1+q_2)^\mu =\left(\sqrt{(q_1+q_2)^2},0,0,0\right).
\end{eqnarray}
The $t$ and the $\bar{t}$ rest frames are then obtained, by boosting (no rotation is involved) into the $t\bar{t}$-ZMF. This is
different from the laboratory frame considered before in Sec.~\ref{sec:cmframe},
where the electron beam is chosen along the $z$ axis, and the $t$ and the $\bar{t}$ rest
frames were constructed by boosting from the lab frame along with a suitable Wigner rotation.

The top quark pair production at $O(\alpha_{\rm em})$ is given by
a direct production with the $\gamma$ and $Z$ exchange: 
\begin{eqnarray}
e^-(p_1,\lambda_1) e^+(p_2,\lambda_2) \overset{\gamma,Z}{\to} t(q_1,s_t) \bar{t}(q_2,s_{\bar{t}})\,.
\label{eq:prod} 
\end{eqnarray} 
The spin four-vectors of the top, $s_t$ and the antitop, $s_{\bar{t}}$ satisfy the usual relations
\begin{equation}
s_t^2 = s_{\bar{t}}^2 = -1\,, \quad\quad  k_1 \cdot s_t = k_2 \cdot s_{\bar{t}} = 0 \,.
\end{equation}
The leading order differential cross section for the $t\bar{t}$ production,
in the presence of longitudinal polarization Eq.~(\ref{eq:fin2}), has the phase space factor
Eq.~(\ref{prod_tt}) and can be written in the spin density matrix representation as
\begin{eqnarray}
&& d\sigma (\lambda_1,\lambda_2,s_t,s_{\bar{t}}) = \frac{3 \beta}{32 \pi s} |\mathcal T|^2 \,,
\nonumber \\
&& 
|\mathcal T|^2 =  \frac{1}{4}{\rm Tr} \left [ \rho \cdot ({\bf 1} + {\bf\hat{s}}_t 
\cdot {\bf\sigma})\otimes ({\bf 1} + {\bf\hat{s}}_{\bar{t}} \cdot {\bf\sigma}) \right ]\,.
\end{eqnarray}
In the above equation, $\rho = \rho^{P(t\bar{t})}$ is the corresponding production spin density matrix 
describing the production of (on-shell) top quark pairs in a specific spin configuration,
while $\mathbf{\hat{s}}_t$ ($\mathbf{\hat{s}}_{\overline{t}}$) is the unit polarization 
vector of the top (antitop) quark in its rest frame and 
$\boldsymbol\sigma = (\sigma_1,\sigma_2,\sigma_3)^T$ is a vector of Pauli matrices.
Conveniently, the most general decomposition of the spin density matrix $\rho$ for the $t\bar{t}$ 
production is of the form
\begin{eqnarray}
\rho = A\, {\bf 1} \otimes  {\bf 1} + B_i^{t} \, \sigma_i  \otimes {\bf 1} +  B_i^{\overline{t}} \, {\bf 1} \otimes \sigma_i + C_{ij} \, \sigma_i \otimes \sigma_j \, , 
\label{eq:rho}
\end{eqnarray}
where the functions $A$, ${B}_i^t$ (${B}_i^{\overline{t}}$) and $C_{ij}$ describe the spin-averaged production
cross section, polarization of top (antitop) quark and the top-antitop spin-spin correlations, respectively.
Using the spin four-vectors defined as
\begin{eqnarray}
s_t^{\mu} &=&  \left ( \frac{ \mathbf{q}_1 \cdot \mathbf {\hat{s}}_t}{m_t}, \mathbf {\hat{s}}_t + \frac{ \mathbf{q}_1 (\mathbf{q}_1 \cdot \hat{\mathbf s}_t)}{ m_t (E_t + m_t)} \right )\,,
\nonumber \\
s_{\overline{t}}^{\mu} &=&  \left (  \frac{ \mathbf{q}_2\cdot \mathbf{\hat{s}}_{\overline{t}}}{m_t}, \mathbf{\hat{s}}_{\overline{t}} + \frac{ \mathbf{q}_2 (\mathbf{q}_2 \cdot \mathbf{\hat{s}}_{\overline{t}})}{ m_t (E_{\bar t} + m_t)} \right )\,,
\end{eqnarray}
the decomposition of the squared scattering amplitude $|\mathcal T|^2$ can be written as
\begin{eqnarray}
|\mathcal T|^2 = a + b^{t}_\mu s_t^{\mu} + b^{\overline{t}}_\mu s_{\overline{t}}^{\mu} + c_{\mu\nu} s_t^{\mu} s_{\overline{t}}^{\nu}\,,
\label{eq:M2}
\end{eqnarray}
and by comparing expressions \eqref{eq:rho} and \eqref{eq:M2} one can extract the functions $A$, ${B}_i^t$ (${B}_i^{\overline{t}}$) and $C_{ij}$.
The functions $B_i^t (B_i^{\bar t})$ and $C_{ij}$ can be further decomposed as 
\begin{eqnarray}
B_i^t &=& b_p^t \hat{p}_i + b_q^t \hat{q}_i \,,\nonumber \\
C_{ij} &=& c_o \delta_{ij} + c_4 \hat{p}_i\hat{p}_j +  c_5 \hat{q}_i\hat{q}_j + c_6 
(\hat{p}_i\hat{q}_j +  \hat{q}_i\hat{p}_j)\,,
\end{eqnarray}
where $\hat{k}$ denotes the unit vector, and we have kept only nonvanishing terms for our case 
\footnote{In the SM the top-quark spin polarization in the normal direction to the production 
plane only exists if one considers QCD radiative corrections or absorptive part of the 
$Z$-propagator. However, since these contributions for the $t\bar{t}$ production at 
linear colliders are extremely small~\cite{GK1,GK2,Groote:2010zf,KornerNEW}
(apart from the threshold region) we do not consider them here.}.

The various top spin observables $\langle {\cal O}_i \rangle$ can then be calculated as
\begin{eqnarray}
 \langle {\cal O}_i (\mathbf{{S}}_t, \mathbf{{S}}_{\overline{t}})\rangle
 = \frac{1}{\sigma} \int d\Phi_{t\bar t} { {\rm Tr} [ \rho \cdot {\cal O}_i (\mathbf{{S}}_t, 
 \mathbf{{S}}_{\overline{t}} ) ]} \,,
 \end{eqnarray}
where $\sigma = \int d\Phi_{t\bar t} {\rm Tr}[\rho]$ is the unpolarized production cross-section,
$d\Phi_{t\bar{t}}$ is the phase space differential and  
$\mathbf{{S}}_t = \boldsymbol{\sigma}/2 \otimes {\bf 1} \, 
( \mathbf{{S}}_{\overline{t}} = {\bf 1} \otimes \boldsymbol{\sigma}/2)$ is the top (antitop) spin operator.  
We consider the following spin observables
\begin{eqnarray}\label{eq:observables}
{\cal O}_1 &=&  \frac{4}{3} \mathbf{{S}}_t \cdot \mathbf{S}_{\overline{t}} \,,
\nonumber \\
{\cal O}_2 &=&  \mathbf{{S}}_t \cdot \mathbf{\hat{a}},~~~~~\bar{\cal O}_2 = \mathbf{{S}}_{\bar t} \cdot \mathbf{\hat{b}} \,,
\nonumber \\
{\cal O}_3 &=& 4 ( \mathbf{{S}}_t \cdot \mathbf{\hat{a}} ) ( \mathbf{S}_{\overline{t}} \cdot \mathbf{\hat{b}} ),\nonumber \\
{\cal O}_4 &=& 4 \left ( (\mathbf{{S}}_t \cdot \mathbf{\hat{p}} ) 
( \mathbf{S}_{\overline{t}} \cdot \mathbf{\hat{q}}) + (\mathbf{{S}}_t \cdot \mathbf{\hat{q}} )
( \mathbf{S}_{\overline{t}} \cdot \mathbf{\hat{p}} ) \right ),
\end{eqnarray}
giving the net spin polarization of the top-antitop system (${\cal O}_1$), polarization of the top (antitop)
quark (${\cal O}_2 (\bar{{\cal O}}_2)$), the top-antitop spin correlation (${\cal O}_3$), with respect to spin quantization axes $\mathbf{\hat a}$ 
and $\mathbf{\hat b}$. The observable ${\cal O}_4$ is an additional top-antitop 
spin correlation with respect to the momentum of the incoming and
the outgoing particles~\cite{BrandUwer}. 

The observable ${\cal O}_1$ can be probed using the opening angle distribution 
($\varphi$), i.e. the angle between the direction of flight of the two (top and antitop) spin analyzers (which are the final particles produced in the top and antitop decays)  
defined in the $t$ and $\bar{t}$ frames, 
respectively, i.e ${\bf\hat{p}}_q \cdot {\bf\hat{p}}_l = \cos\varphi$, 
\begin{eqnarray}\label{eq:dobservable}
\frac{1}{\sigma} \frac{d \sigma}{d \cos \varphi} &=& \frac{1}{2} \left (  1 - D  \cos\varphi \right ),
\end{eqnarray}
and
\begin{equation}
D = \braket{\mathcal O_1} \kappa_f \kappa_{\bar{f}} 
\end{equation}
where $\kappa_f(\kappa_{\bar{f}})$ are the top, antitop spin analyzers considered here. The spin analyzer 
for the FCNC top-Higgs decays can be either a direct $t$-quark daughter, i.e. $H$ or $c/u$-quark, 
or $H$ decay products like $b$ or $\bar{b}$ in $b\bar{b}$ decay, or $\tau^+(\tau^-)$ in $H \to \tau^+\tau^-$ decay, or jets. On the other hand, 
the spin analyzer for $\bar{t}$ are $W^-$ or $\bar{b}$, or a $W^-$ decay products $l^-, \bar \nu$ or jets. 
We consider the $q=c/u$ quark from the top and the $l^-$ from the antitop as spin analyzers in this work.
The spin analyzers are calculated from the one-particle decay density matrices given as 
\begin{eqnarray}
\rho^{t \to f(\bar{t} \to \bar{f})}_{\alpha\alpha'} &=& \Gamma^{t \to f(\bar{t} \to \bar{f})}
\left[\frac{1}{2} \left({\bf 1} + \kappa_{f(\bar{f})} 
{\bf\hat{p}}_{f(\bar{f})}\cdot \bf{\sigma} \right)\right]_{\alpha\alpha'}. 
% \rho^{\bar{t} \to a_2} &=& \frac{\Gamma^{(\bar{t} \to a_2)}}{2} \left({\bf 1} + \kappa_{a_2}
% {\bf\hat{p}}_{a_2}\cdot \bf{\sigma} \right)
\end{eqnarray}
where $\alpha,\alpha'$ denote the $t$-quark spin orientations, ${\bf\hat{p}}_{f}$ and ${\bf\hat{p}}_{\bar{f}}$ are the directions of flight 
of the final particles $f$ and $\bar {f}$ in the rest frame of the top and the antitop quarks respectively. 
The values of various $\kappa_{f(\bar f)}$ for SM top (antitop) decays are presently known at NLO in QCD
and can be found in~\cite{Brandenburg:2002xr}. The top quark polarization matrix can be also written as 
 \begin{equation}
\rho^{t \to f}_{\alpha\alpha'} = \frac{1}{2} 
\begin{pmatrix}
1 + \kappa_f \cos\theta^{top}_f  & \kappa_f \sin \theta^{top}_f e^{i \phi^{top}_f} \\
\kappa_f \sin \theta^{top}_f e^{-i \phi^{top}_f}  &1 - \kappa_f \cos\theta^{top}_f 
                                               \end{pmatrix}_{\alpha\alpha'},
 \end{equation}
 and similarly for the antitop spin matrix $\rho^{\bar{t} \to \bar{f}}$.
The top spin analyzing power of $q$ ($\kappa_{q}$) from the $t \to H q$ decay
can be calculated from Eq.~(\ref{eq_me_topch_rest}), in Appendix~\ref{sec:appen_decaytcH}, 
%  
%  
% The polarization degree $\kappa_{f}$ of the top is then equal to
% \begin{equation}
%  \kappa_{f} = \frac{\rho^{t\to f}_{\uparrow \uparrow}-
%  \rho^{t \to f}_{\downarrow \downarrow}}{\rho^{t\to f}_{\uparrow \uparrow}+\rho^{t\to f}_{\downarrow \downarrow}}.
% \end{equation}
% A similar expression holds for the antitop, $\kappa_{\bar{f}}$. The value of $\kappa_q$ from the $t \to H q$ decay
% can be calculated from Eq.~(\ref{eq_me_topch_rest}), in Appendix~\ref{sec:appen_decaytcH}, 
\begin{eqnarray}
\kappa_q &=& \frac{|g_{qt}|^2-|g_{tq}|^2}{|g_{qt}|^2+|g_{tq}|^2} \,.\label{eq:kappa1}
\end{eqnarray}
 Similarly, 
the spin analyzing power for the $b$ quark ($\kappa_b$), from the top decay to $W^+b$  can be obtained from 
Eqs.~(\ref{eq_me_topWB_rest}), in Appendix~\ref{sec:appen_decaytWb}, 
\begin{eqnarray}
 \kappa_b &=& \frac{m_t^2-2m_W^2}{m_t^2+2m_W^2} \,. \label{eq:kappa2}
\end{eqnarray}
Leptons emitted from the antitop decay, due to the $V-A$ interactions
are the perfect top spin analyzers (Eqs.~(\ref{eq_me_top}), in Appendix~\ref{sec:appen_decaytWb}) with 
\begin{equation}
\kappa_{\bar{f}} = \kappa_l =1,
\end{equation} 
at LO QCD ($\alpha_s$ corrections are negligible \cite{Brandenburg:2002xr}),   
with their flight directions being 100\% correlated with the directions of the top spin.
It is clear from Eq.~(\ref{eq:kappa1}) that with $|g_{qt}|^2 \simeq |g_{tq}|^2$, the spin information of the top is lost ($\kappa_q \approx$ 0). However in the presence or dominance of only one of the coupling,
the emitted quark acts as a perfect spin analyzer ($\kappa_q \approx 1$).  

The top (antitop)-quark polarization and spin-spin correlations can be measured using the double differential 
angular distribution of the top and antitop quark decay products:
\begin{eqnarray}
\frac{1}{\sigma} \frac{d^2 \sigma}{d \cos\theta_f d\cos\theta_{\bar f}} = \frac{1}{4} \left ( 1 + B_t 
 \cos\theta_f + B_{\bar{t}}  \cos\theta_{\bar f} - C \cos\theta_f \cos\theta_{\bar f} \right )\,,
\label{eq:diffsigma}
\end{eqnarray}
where $\theta_f (\theta_{\bar f})$ is the angle between the direction of the top (antitop) 
spin analyzer $f,\,(\bar f)$ in the $t$ $(\bar{t})$ rest frame and the 
$\hat{\bf a}$ ($\hat{\bf  b}$) direction in the $t\bar{t}$-ZMF, c.f.~\cite{Bernreuther:2004jv}.  
Comparing Eq.~(\ref{eq:diffsigma}), with Eq.~(\ref{eq:observables}), we have
\begin{eqnarray}
B_t  &=& \braket{\mathcal O_2} \kappa_{f} \,,\quad 
\quad B_{\bar t}  = \braket{\bar {\mathcal O}_2} \kappa_{\bar f}\,,
\nonumber \\
C  &=& \braket{\mathcal O_3} \kappa_{f} \kappa_{\bar f}\,.
\end{eqnarray}
where ${\cal O}_2$ and $\bar{\cal O}_2$ are related to the top, antitop spin polarization 
coefficients $B_t$ and $B_{\bar t}$. Since there is no CP violation in our case, we consider
$B \equiv B_t = \mp B_{\bar t}$ for $\hat{\bf a} = \pm \hat{\bf b}$ . 
This limit is a good approximation for the charged leptons 
from $W$ decays~\cite{Brandenburg:2002xr}. 
The spin observable ${\cal O}_3$ is also related to the spin correlation function $C_{ij}$ 
in Eq.~(\ref{eq:rho}), 
\begin{eqnarray}
\braket{ {\cal O}_3 } = \frac{ \sigma_{t \bar{t}} (\uparrow \uparrow) + \sigma_{t \bar{t}} (\downarrow \downarrow) - \sigma_{t \bar{t}} (\uparrow \downarrow)- 
\sigma_{t \bar{t}} (\downarrow \uparrow)}
{\sigma_{t \bar{t}} (\uparrow \uparrow) + \sigma_{t \bar{t}} (\downarrow \downarrow) + \sigma_{t \bar{t}} (\uparrow \downarrow)+ \sigma_{t \bar{t}} (\downarrow \uparrow)}\,,
\end{eqnarray}
where the arrows refer to the up and down spin orientations of the top and the antitop 
quark with  respect to the $\mathbf{\hat a}$ and $\mathbf{\hat b}$ quantization axes,
respectively. 
\\
Also ${\cal O}_4$ gets corrected by $\kappa_{f} \kappa_{\bar f}$ depending on the final particles measured from the $t$ and $\bar{t}$ decays. 

The arbitrary unit vectors $\bf{\hat a}$ and $\bf{\hat b}$ specify different
spin quantization axes which can be chosen to maximize/minimize the desired polarization 
and the correlation effects. We work with the following choices:
\begin{align}
\hat{\bf a} &= - \hat{\bf b} = \hat{\bf q}\,,  && ({\rm ``helicity" \; basis})\,,&&
\nonumber \\
\hat{\bf a} &= \hat{\bf b} = \hat{\bf p}\,,  && ({\rm ``beamline" \; basis})\,,&&
\nonumber \\
\hat{\bf a} &= \hat{\bf b} = \hat{\bf d}_{\bf X}\,,  && ({\rm ``off-diagonal" \; basis\, (specific \, for\,some \,model\, X)})\,,&&
\nonumber \\
\hat{\bf a} &= \hat{\bf b} = \hat{\bf e}_{\bf X}  &&({\rm ``minimal" \; basis\, (specific \, for\,some \,model\, X)})
\label{eq:axes}
\end{align}
where $\hat{\bf p}$ is the direction of the incoming beam and $\hat{\bf q} = \hat{\bf q}_1$ is
the direction of the outgoing top quark, both in the $t\bar t$ center of mass frame. The 
off-diagonal basis \cite{Parke:1996pr} is the one, where the top spins are $100\%$ correlated and
is given by quantizing the spins along the axis $ \hat {\bf d}_{\rm SM}$ determined as
\begin{eqnarray}
\hat{\bf d}_{\rm SM} = \hat{\bf d}_{\rm SM}^{\rm max} =
\frac{ - \hat{\bf p}  + ( 1 - \gamma) z \; \hat{\bf q}_1}{\sqrt{ 1 - (1 - \gamma^2) z^2}}\,,
\label{eq:dSM}
\end{eqnarray}
where $z = \hat{\bf p} \cdot \hat{\bf q}_1 = \cos \theta$ and 
$\gamma = E_t/m_t = 1/\sqrt{1 - \beta^2}$ and which interpolates between the beamline 
basis at the threshold ($\gamma \to 1$) and the helicity basis for
ultrarelativistic energies ($\gamma \to \infty$). We would like to point out here that this
off-diagonal basis $\hat{\bf d}_{\rm SM}$ is specific to the SM $t\bar{t}$ production, but a
general procedure for finding such an off-diagonal basis is given 
in~\cite{Mahlon:1997uc, Uwer:2004vp}. The idea is to determine the maximal eigenvalue
of the matrix function $C_{ij}$ in Eq.~(\ref{eq:rho}) and the corresponding eigenvector,
which provides the off-diagonal quantization axis $\hat {\bf d}_X$, for any model X 
\cite{KamenikMelic}.

Here we introduce the complementary basis to the ``off-diagonal'' one, $\bf{\hat{e}}_{\rm SM}$, 
where the eigenvector corresponds to the minimal eigenvalue of $C_{ij}$  in the
SM quark-antiquark production. The correlation of the top-antitop spins
in this basis is minimal. This axis could be useful in the 
new model searches since the minimization of the top-antitop correlations in the SM 
can, in principle, enhance the non-SM physics. The `minimal basis' is defined by the axis  
\begin{eqnarray}
{\bf\hat{e}}_{\rm SM} = \hat{\bf e}_{\rm SM}^{\rm min} = 
\frac{ - \gamma z  {\bf\hat{p}}  +  (1 - ( 1 - \gamma^2) z^2 )\; 
{\bf\hat{q}_1}}{\sqrt{ (1 - z^2) (1 - (1 - \gamma^2) z^2)}}.
\label{eq:eSM}
\end{eqnarray}
\begin{figure}
\hspace{3.5cm}
\includegraphics[width=6cm, height=4.5cm]{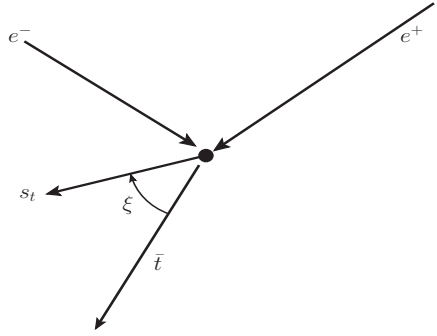}
\caption{The top quark spin vector $s_t$ in the $t\bar{t}$ production in $t$ rest frame, with
the direction of $s_t$ given by an angle $\xi$. The angle $\xi$ is measured in the 
clockwise direction from the $\bar{t}$ momenta.}
\label{fig:spin}
\end{figure}
The `off-diagonal' and the `minimal' basis define the angle $\xi$ between the top-quark
spin vector and the antitop direction in the top-quark rest frame~\cite{Mahlon:2010gw}, 
shown in Fig.~\ref{fig:spin}, 
\begin{eqnarray}
\tan \xi^{\rm off(=max)} &=& \frac{\tan \theta_t }{\gamma} \,, \quad
\tan \xi^{\rm min} = \frac{\gamma }{\tan \theta_t} \,,
\end{eqnarray}
or 
\begin{eqnarray}
 \cos^2 \xi^{\rm min} + \cos^2 \xi^{\rm off(=max)} = 1 \,,
\end{eqnarray} 
as expected. As already stated, such axes which minimize or maximize spin correlations can be constructed for any model. 

The analytical form of the observables defined in Eq.~(\ref{eq:observables}), 
is listed in appendix~\ref{sec:appen_spin} for the SM $t\bar{t}$ production in the presence
of longitudinal polarization. The observables ${\cal O}_i$, ($i$ = 1,2,3,4) are then 
multiplied with the appropriate $\kappa$ factors. The QCD radiative corrections for all 
the top spin observables considered here are calculated in \cite{BrandUwer} and it is shown 
to be small. Also recently it has been shown that the $O(\alpha_S)$ corrections to the 
maximal spin-spin correlations in the off-diagonal basis are negligible \cite{KornerNEW}. 
Therefore we neglect them all in our calculations. 

Next, we present the results for spin correlations
and top (antitop)-quark polarizations in the helicity basis ($C_{\rm hel}$, $B_{\rm hel}$),
beamline basis ($C_{\rm beam}$, $B_{\rm beam}$), off-diagonal ($C_{\rm off}$, $B_{\rm off}$)
and the minimally polarized basis ($C_{\rm min}$, $B_{\rm min}$), as 
defined by Eqs.~\eqref{eq:axes}, \eqref{eq:dSM} and~\eqref{eq:eSM} respectively, and check
for their sensitivity to the initial beam polarization. These results are presented in the 
absence of cuts, realistic cuts severely distort the non-zero coefficients of 
Eq.~(\ref{eq:dobservable}) and Eq.~(\ref{eq:diffsigma}). The observable ${\cal O}_1$ as seen
from Eq.~(\ref{eq:O1}), is equal to 1 and is therefore independent of beam polarization. 
However, it is dependent on the value of $\kappa_f$. 

In Table~\ref{tab:O1-4} we present the values of the different spin observables
in the different spin basis considered here, in the presence of beam polarizations. We have
considered the case, when the antitop is decaying to lepton ($\kappa_{\bar{f}}$ =1), 
$\kappa_f = \kappa_q$, Eq.~(\ref{eq:kappa1}) for the  
FCNC top decay, and $\kappa_f = \kappa_b$, Eq.~(\ref{eq:kappa2}) for the top decaying to $W^+b$.
\begin{table}[htb]
\begin{center}
\begin{tabular}{|c|c|c|c|c|} \hline
Observables &Basis &$P^L_{e^-} = 0, P^L_{e^+} = 0$ &$P^L_{e^-} = 0.8, P^L_{e^+} = -0.3$
&$P^L_{e^-} = -0.8, P^L_{e^+} =0.3$ \\ \hline
${\cal O}_1$ &      &0.333$\kappa_f$  &0.333$\kappa_f$  &0.333$\kappa_f$    \\ \hline
          &hel    &$-$0.076$\kappa_f$  & 0.247$\kappa_f$  &$-$0.239$\kappa_f$    \\ 
          &beam    &$-$0.174$\kappa_f$  & 0.344$\kappa_f$  &$-$0.436$\kappa_f$    \\ 
${\cal O}_2$ &off   &0.176$\kappa_f$  &$-$0.351$\kappa_f$  & 0.443$\kappa_f$    \\ 
          &min    & 0.04$\kappa_f$  &$-$0.131$\kappa_f$  & 0.127$\kappa_f$    \\ \hline
          &hel    & $-$0.654$\kappa_f$  & $-$0.666$\kappa_f$  & $-$0.648$\kappa_f$    \\ 
          &beam    & 0.881$\kappa_f$  &0.852 $\kappa_f$  &0.897$\kappa_f$    \\ 
${\cal O}_3$ &off   &0.911$\kappa_f$  & 0.886$\kappa_f$  &0.924$\kappa_f$    \\ 
          &min    & 0.224$\kappa_f$  &0.229 $\kappa_f$  &0.222$\kappa_f$    \\ \hline                                     
${\cal O}_4$ &      & 0.546$\kappa_f$  & 0.612$\kappa_f$  &0.512$\kappa_f$    \\ \hline                                    
\end{tabular}
\caption{The value of the spin observables in different bases, with different choices  of initial beam polarization. $\kappa_f = \kappa_q$ for FCNC $t$-decays and $\kappa_f = \kappa_b$ for $t \to W^+ b$.}
\label{tab:O1-4}
\end{center}
\end{table} 
We note that the top (antitop) spin polarizations are quite sensitive to the beam polarization, while 
this is not the case for the spin-spin correlations ${\cal O}_3,{\cal O}_4$ where the influence of the beam 
polarizations gets diluted, see Eqs.~(\ref{eq:O3h}-\ref{eq:O4}). 
%,~\ref{eq:O3b},~\ref{eq:O3o},~\ref{eq:O3m},~\ref{eq:O4}.
Also note that all observables are proportional to $\kappa_f = \kappa_q$ and  will be equal to zero if $g_{tq}$ and $g_{qt}$ are equal. 
%%%%%%%%%%%%%%%%%%%%%%%%%%%%%%%%%%%%%%%%%%%%%%%%%%%%%%%%%%%%%%%%%%%%%%%%%%%%
%%%%%%%%%%%%%%%%%%%%%%%%%%%%%%%%%%%%%%%%%%%%%%%%%%%%%%%%%%%%%%%%%%%%%%%%%%%%
%%%%%%%%%%%%%%%%%%%%%%%%%%%%%%%%%%%%%%%%%%%%%%%%%%%%%%%%%%%%%%%%%%%%%%%%%%%%
%%%%%%%%%%%%%%%%%%%%%%%%%%%%%%%%%%%%%%%%%%%%%%%%%%%%%%%%%%%%%%%%%%%%%%%%%%%%
%%%%%%%%%%%%%%%%%%%%%%%%%%%%%%%%%%%%%%%%%%%%%%%%%%%%%%%%%%%%%%%%%%%%%%%%%%%%
%%%%%%%%%%%%%%%%%%%%%%%%%%%%%%%%%%%%%%%%%%%%%%%%%%%%%%%%%%%%%%%%%%%%%%%%%%%%
\section{Numerical analysis of the FCNC $g_{tq},~g_{qt}$ couplings at the ILC}\label{sec:numericalstudy}

In this section we perform a detailed numerical simulation of the FCNC interactions in the $t \to qH$ 
decay at the ILC. As before, the process we consider is the top pair production, 
with the top decaying to $qH$, the antitop decaying to $W^-\bar{b}$ with the $W^-$ decaying
leptonically and subsequently the Higgs decaying to a $b\bar{b}$ pair.
The main background for the process under study comes from the $t\bar{t}$ pair production, with 
one of the top decaying hadronically and the other decaying to a lepton, $\nu$ and a $b$ quark.
We have performed our calculations, by first generating the Universal Feynrules Output (UFO) model 
file using FeynRules 2.3~\cite{Alloul:2013bka}, including
the effective interaction, defined in Eq.~(\ref{eq:tqhA}). The UFO file is then implemented in
MadGraph 5 v2.4.2~\cite{Alwall:2011uj}, for Monte Carlo simulation. We also employ
Pythia 8~\cite{Sjostrand:2014zea} for parton showering and hadronization along with 
Fastjet-3.2.0~\cite{Cacciari:2011ma} for the jet formation.
The cross section of the signal and the background, at $\sqrt{s}$ = 500 GeV, before the
application of the event selection criteria is listed in Table~\ref{tab:allowedcs}. 
\begin{table}[htb]
\begin{center}
\begin{tabular}{|c|c|c|c|} \hline
&$\sigma$(fb)&$\sigma$(fb)&$\sigma$(fb) \\
$e^-e^+\rightarrow t\bar{t}$ &$P^L_{e^-} = 0, P^L_{e^+} = 0$  & $P^L_{e^-} = -0.8, P^L_{e^+} = 0.3$ 
& $P^L_{e^-} = 0.8, P^L_{e^+} = -0.3$  \\ 
 \hline
% &&& \\
{\rm signal:}&&& \\$t\rightarrow q b\bar{b},
\bar{t}\rightarrow  l^- \bar{\nu}_l \bar{b}$ &$73.4(|g_{tq}|^2+|g_{qt}|^2)$ 
&$120.5(|g_{tq}|^2+|g_{qt}|^2)$ &$62(|g_{tq}|^2+|g_{qt}|^2)$  \\ \hline
% &&& \\
{\rm background:}&&&\\
$t\rightarrow q_1 q_2 b,
\bar{t}\rightarrow  l^- \bar{\nu}_l \bar{b}$  &74.5 &124.7 &58.9 \\ \hline
\end{tabular}
\caption{The production cross section of the signal and the background at $\sqrt{s}$ = 500 GeV. The results
are presented for both the polarized and the unpolarized beams.}
\label{tab:allowedcs}
\end{center}
\end{table}
%The $t\bar{t}$ acts as a dominant background for the process under study. 

We now describe in details 
the different cuts and conditions considered for our analysis. Since the top
from the $tqH$ final state decays to $Wb$, the lepton from the $W$, tends to be energetic
and isolated. Therefore firstly the events 
with one isolated lepton are selected, through the lepton isolation cut. An isolated lepton is identified, 
by demanding that the scalar sum of the energy of all the stable
particles within the cone of $\Delta R = \sqrt{\Delta \eta^2 +\Delta\phi^2}\leq 0.2$ about 
the lepton is less than  $\sqrt{6(E_l-15)}$~\cite{Yonamine:2011jg}, where
$E_l$ is the energy of the lepton. Furthermore, the transverse momenta of the 
leptons are assumed as $p_T>$10 GeV. The events with more than one isolated lepton are discarded.
The remaining stable visible particles of the event, are then clustered into four jets using the 
inbuilt $k_t$ algorithm in FastJet for $e^-e^+$ collisions, which is similar to the Durham algorithm.
The reconstructed jets and the isolated lepton are combined to form the intermediate heavy states. 
The three jets with the highest $b$ tagging probability are considered as the $b$ jets. A jet 
is tagged as a $b$ jet if it has a $b$ parton within a cone of $\Delta R <$ 0.4 with the jet axis.
A tagging efficiency of 80\%~\cite{Asner:2013psa} is further incorporated. The jets are checked 
for isolation and are expected to have $p_T>$ 20 GeV.
The momentum of the neutrino is calculated by summing over all the visible momenta
and the energy of the neutrino is assigned the magnitude of its momenta vector. The isolated lepton and the neutrino reconstructs the leptonically decaying $W$ boson. 

There will be three $b$ tagged jets and a non $b$ jet in the final state and therefore
three possible combinations to reconstruct the Higgs mass from the $b$ tagged jets. Additionally
one of this pair of $b$ jets reconstructing the Higgs mass, along with the the non $b$ jet should
give an invariant mass close to $m_t$. We choose the combination of the jets,
which minimizes the quantity $|m_{b_ib_j}-m_H|^2 +|m_{b_ib_jQ}-m_t|^2$, with 
$i,j$ taking values for various combinations of the $b$ jets and $Q$ is the non-$b$ jet. 
The reconstructed Higgs mass is given by $m_{b_ib_j}$, and the reconstructed top
mass is denoted by $m_{b_ib_jQ}$. In order to account for the detector resolution, 
we have smeared the leptons and the jets using the following parametrization. The
jet energies are smeared~\cite{BrauJames:2007aa} with the different 
contributions being added in quadrature,
\begin{eqnarray}
 \frac{\sigma(E_{jet})}{E_{jet}}&=& \frac{0.4}{\sqrt{E_{jet}}} \oplus 2.5\% \,.
\end{eqnarray}
The momentum of the lepton is smeared as a function of the momentum and the angle $\cos\theta$ 
of the emitted leptons~\cite{Li:2010ww}
\begin{eqnarray}
 \frac{\sigma(P_l)}{P_l^2} =\left(\begin{array}{cc}a_1\oplus\frac{b_1}{P_l},&|\cos\theta_l|< 0.78 \\
                \left(a_2 \oplus \frac{b_2}{P_l}\right)  \left(\frac{1}{\sin(1-|\cos\theta_l|)}
                \right) & |\cos\theta_l|> 0.78 
               \end{array}\right) \,,
\end{eqnarray}
with
\begin{eqnarray}
(a_1,~b_1) &=& 2.08\times10^{-5}~({\rm{1/GeV}}),~~8.86\times10^{-4}, \nonumber \\
(a_2,~b_2) &=& 3.16\times10^{-6}~({\rm{1/GeV}}),~~2.45\times10^{-4}. 
\end{eqnarray}

\begin{figure}[htb]
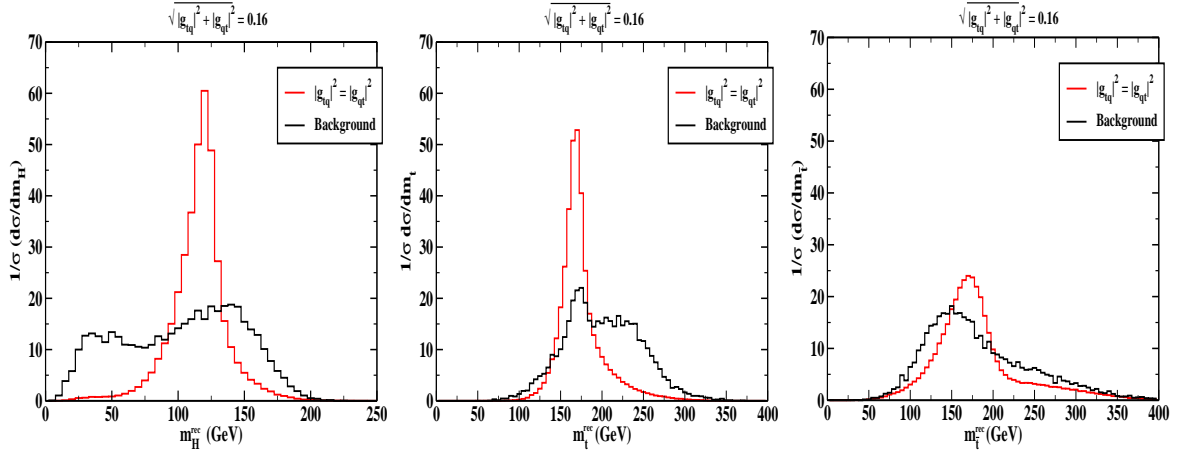

\vspace{1cm}
$\begin{array}{ccc}
 \includegraphics[width=5 cm, height= 6cm]{mHiggs.eps} &
  \includegraphics[width=5 cm, height= 6cm]{mass_t.eps} &
    \includegraphics[width=5 cm, height= 6cm]{mass_tbar.eps}
    \end{array}$
 \caption{The reconstructed masses of the Higgs, $t$-quark and $\bar{t}$, for the signal 
and the $t\bar{t}$ background, at $\sqrt{s}$ =  500 GeV, with ${\cal L}$ = 500 fb$^{-1}$
and unpolarized beams. For the 
signal we have considered Case 1 from Sec.~\ref{sec:cmf_asymmetries}  with $\sqrt{|g_{tq}|^2+|g_{qt}|^2}$ = 0.16.}
  \label{fig:m_recons}
\end{figure}

\begin{figure}
\centering
\begin{minipage}{0.45\textwidth}
\centering
\includegraphics[width=6.5cm, height=6cm]{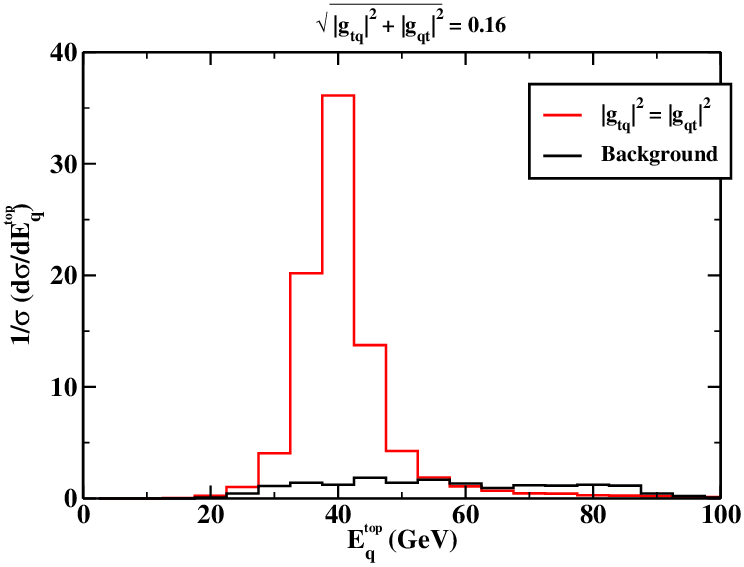}
\caption{The energy distribution of the non-$b$ jet ($t\rightarrow qH$) in the rest frame of the top, 
at $\sqrt{s}$ = 500 GeV, with unpolarized beams and ${\cal L}$ = 500 fb$^{-1}$.}
\label{fig:energy_rf}
\end{minipage}
\hspace{0.5cm}
\begin{minipage}{0.45\textwidth}
\centering
\includegraphics[width=6.5cm, height=6cm]{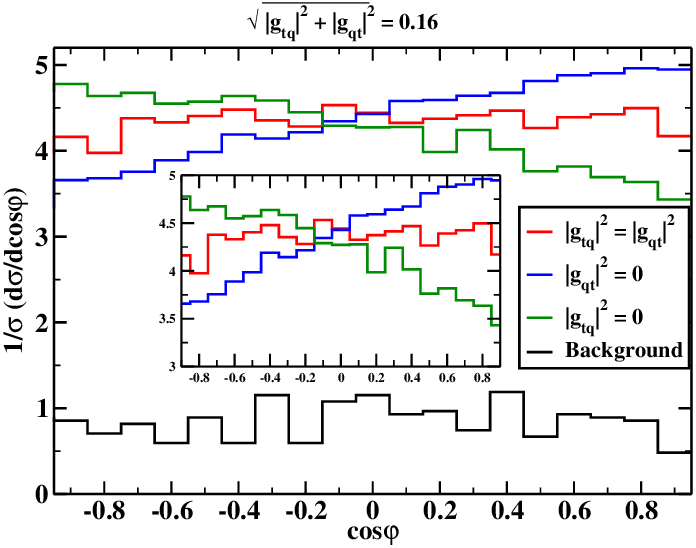}
\caption{The opening angle distribution, Eq.~(\ref{eq:dobservable}) between the
direction of the lepton (from $\bar{t}\rightarrow l^-\bar{\nu} \bar{b}$)  and the non-$b$ jet (from $t\rightarrow qH$), in the $t$ and $\bar{t}$ rest frame. }
\label{fig:lepjet}
\end{minipage}
\end{figure}

We plot in Fig.~\ref{fig:m_recons}, the reconstructed Higgs, $t$ and the $\bar{t}$ masses. The 
Higgs mass is reconstructed as $m_H^2 = (p_b +p_{\bar{b}})^2$, whereas the top 
the antitop masses are calculated as 
$m_t^2 = (p_b + p_{\bar{b}} + p_{non-b})^2$, 
$m_{\bar{t}}^2 = (p_{l^-}+ p_{\bar{\nu}}+ p_{\bar{b}})^2$. 
The plots for the signal are constructed 
taking into account the current stringent LHC constraint on the FCNC couplings, $\sqrt{|g_{tq}|^2+|g_{qt}|^2}$ = 0.16. We have shown the results for Case 1, discussed
in Sec.~\ref{sec:cmf_asymmetries}, as the reconstructed mass will be the same for all three cases.
We note that since we have not done a real detector analysis, the mass reconstruction of the $W$ boson is poor
in our case, due to the presence of missing energy. Therefore a loose cut on $m_W$ is applied for our analysis.
It is clear from Fig.~\ref{fig:m_recons}, that the cut imposed on the reconstructed $m_t$ and 
$m_{\bar{t}}$ should be different. The reconstructed mass of $\bar{t}$ is broad, due
to the presence of the missing energy from the $W$ decay.  We have applied the same kinematic cut to the mass
of the top and the antitop for the sake of simplicity. The implementation of these cuts, eliminates the 
$Wb\bar{b}jj$ and $Zb\bar{b}jj$ backgrounds.
The kinematical cuts, which are imposed on the various reconstructed masses are summarized below:
\begin{itemize}
\item 115 $\leq m_H$ (GeV) $\leq$ 135,~~~~160 $\leq m_t$ (GeV) $\leq$ 188~~~~30 $\leq m_W$ (GeV) $\leq$ 100
\end{itemize}
Additional cuts can be applied, on the energy of the emitted quark in the top rest frame~\cite{Han:2001ap}, so as 
to increase the signal to background ratio. The energy of the emitted quark, as a result of the
two body decay of the top is 
\begin{equation}
 E_{q}^{top}=\frac{m_t}{2}\left(1-\frac{m_H^2}{m_t^2}\right),
\end{equation}
and is peaked around 42 GeV, for a Higgs mass of 125 GeV. The jet from the background, which will fake the $q$
jet, will have a more spread out energy. We do not apply this cut, as the application of the above cuts 
already lead to a much reduced background. The energy distribution of both the signal and the background
are shown in Fig.~\ref{fig:energy_rf}.

\begin{figure}[htb]
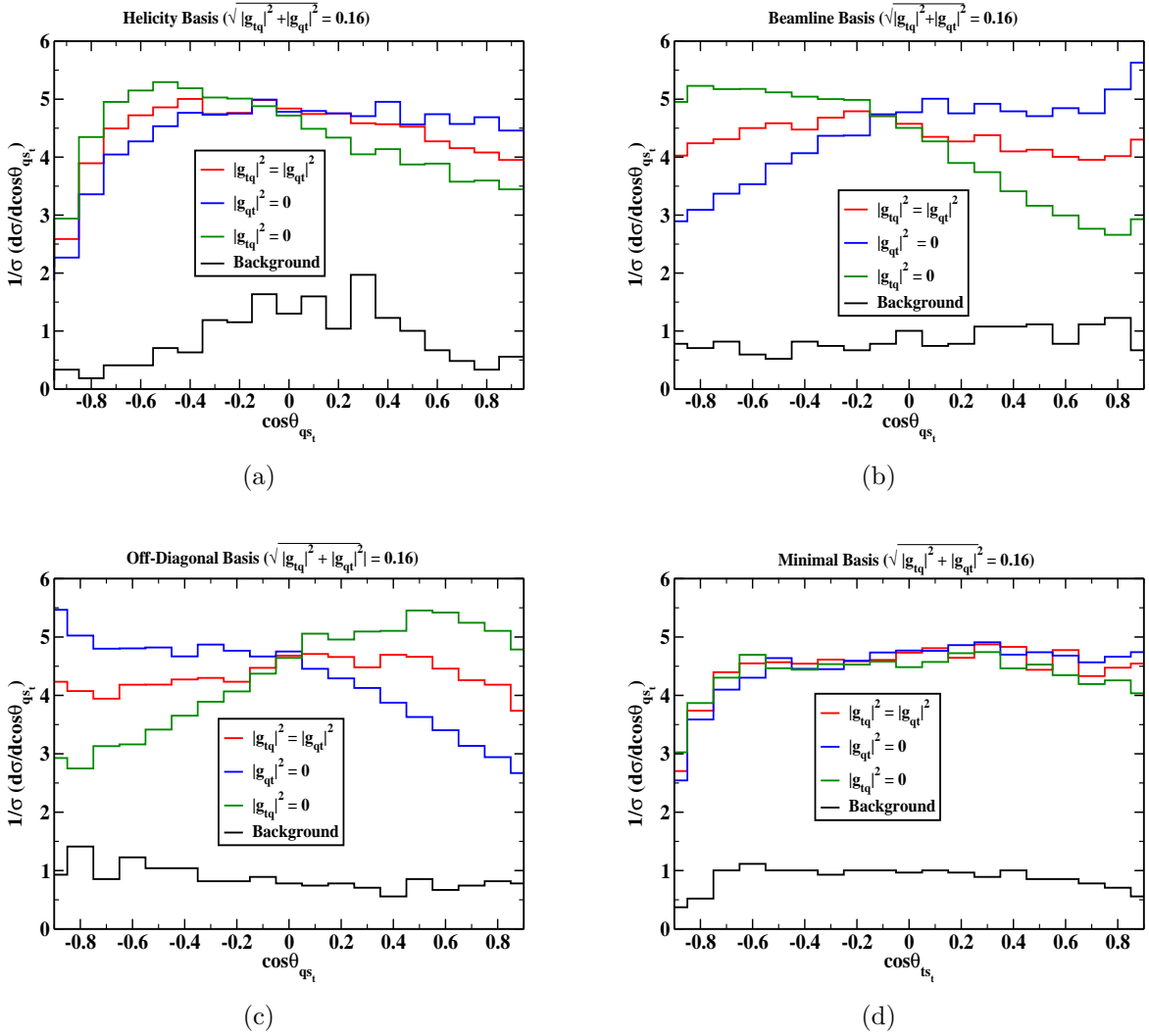

\vspace{0.4cm}
  \begin{subfigure}{0.45\linewidth}
  \centering
    \includegraphics[width=7cm, height=6cm]{jettop_hb.eps}
     \caption{}
    \label{fig:hb}
  \end{subfigure}
   \hspace{1.0cm}
  \begin{subfigure}{0.45\linewidth}
  \centering
    \includegraphics[width=7cm, height=6cm]{jettop_bb.eps}
     \caption{}
    \label{fig:bb}
  \end{subfigure}
    \begin{subfigure}{0.45\linewidth}
\centering
\vspace{0.6cm}
  \includegraphics[width=7cm, height=6cm]{jettop_odb.eps}
     \caption{}
    \label{fig:odb}
  \end{subfigure}
  \hspace{1.3cm}
  \begin{subfigure}{0.45\linewidth}
  \vspace{0.6cm}
\centering  
    \includegraphics[width=7cm, height=6cm]{jettop_mb.eps}
     \caption{}
    \label{fig:mb}
  \end{subfigure}
  \caption{The distribution $1/\sigma (d\sigma/d\cos\theta_{qs_t})$, with unpolarized beams
  at $\sqrt{s}$ = 500 GeV and $\cal L$ = 500 fb$^{-1}$, where $\theta_{qs_t}$ is the 
angle between the direction of the top spin analyzer (non-$b$ jet from $t\rightarrow q H$) in 
$t$ rest frame and the spin quantization axis of
the top ($s_t$) in the $t\bar{t}$-ZMF. The different spin quantization axes considered are discussed
in Eq.~(\ref{eq:axes}).}
\label{fig:basis}
\end{figure}

\begin{figure}
\centering
\begin{minipage}{0.45\textwidth}
\centering
\includegraphics[width=6.5cm, height=6cm]{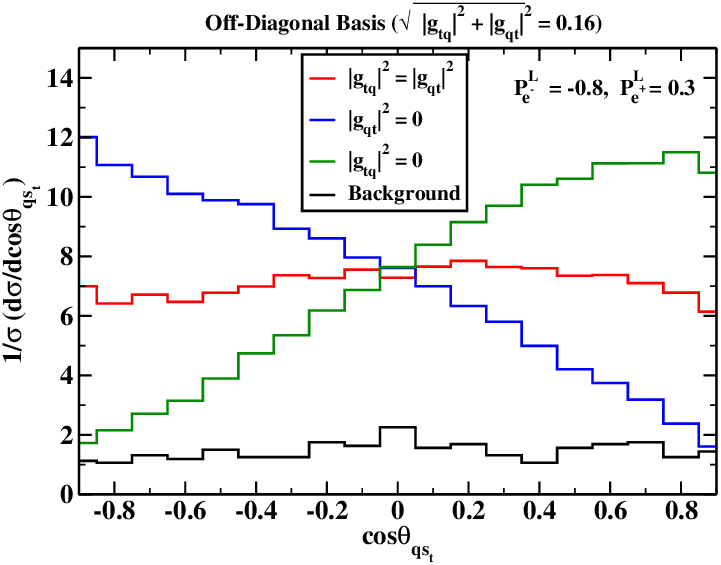}
\caption{The normalized $1/\sigma (d\sigma/d\cos\theta_{qs_t})$ distribution
(the definitions are same as in Fig.~\ref{fig:basis}) at $\sqrt{s}$ = 500 GeV,
with polarized beams and ${\cal L}$ = 500 fb$^{-1}$.}
\label{fig:offDB}
\end{minipage}
\hspace{0.5cm}
\begin{minipage}{0.45\textwidth}
\centering
\includegraphics[width=6.5cm, height=6cm]{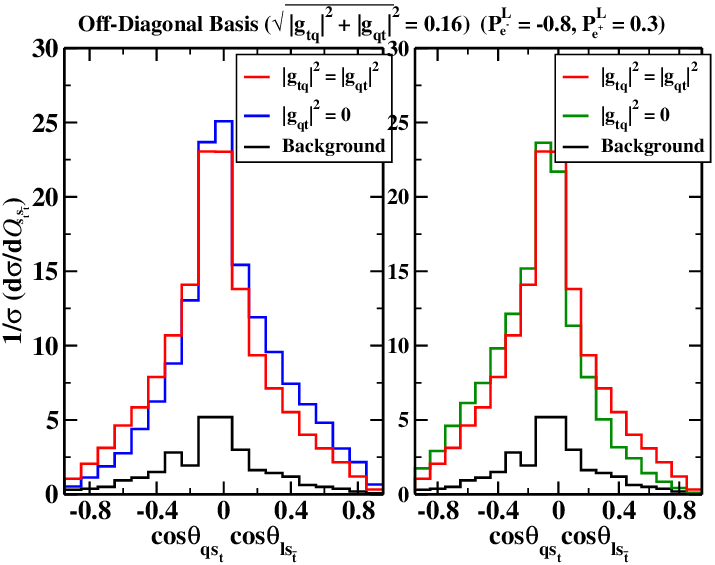}
\caption{The normalized distribution of the product 
$\cos\theta_{qs_t} \cos\theta_{ls_{\bar{t}}}$ , 
($\theta_{qs_t} = \angle (\hat{\bf p}_q,\hat{\bf a}), 
\theta_{qs_{\bar{t}}} = \angle (\hat{\bf p}_l,\hat{\bf b})$),
using the off-diagonal basis, at $\sqrt{s}$ = 500 GeV, with 
polarized beams and ${\cal L}$ = 500 fb$^{-1}$. }
\label{fig:coscos}
\end{minipage}
\end{figure}

Further on, we concentrate on the observables which will be sensitive to the chiral nature of the FCNC interactions. 
One of them is the polar angle distribution of the non-$b$ jet, which was earlier shown in Fig.~\ref{fig:dist_pol}.
The effect of the individual chiral couplings is more evident with a suitable choice of initial 
longitudinal beam polarization.  The various distributions which we consider here 
are all calculated
in the $t\bar{t}-$ZMF. The decay products, which act as spin analyzers for our case are 
the non-$b$ jet ($q$) from the decay $t\rightarrow q H$ and the lepton ($l^-$)from the decay 
$\bar{t} \rightarrow l^-\bar{\nu} \bar{b}$. All the distribution plots are given with the 
number of surviving events, for $\mathcal{L}$ = 500 fb$^{-1}$.
We plot the opening angle distribution $1/\sigma (d\sigma/d\cos\varphi)$ (Eq.~(\ref{eq:dobservable}))
in Fig.~\ref{fig:lepjet}, which is
sensitive to the top and the antitop spin analyzers.  The distribution is
flat for Case 1, when $|g_{tq}|^2 = |g_{qt}|^2$, leading 
to $\kappa_q = 0$. It peaks in the forward direction in the presence of $|g_{tq}|^2$, and in the backward
direction for $|g_{qt}|^2$ (clearly seen in the inset of Fig.~\ref{fig:lepjet}). 
The top spin
%correlations being dependent on the product of the top decay spin analyzers are 
is considered
in the normalized distribution $1/\sigma (d\sigma/d\cos\theta_{qs_t})$, where 
$\theta_{qs_t}$ is the angle between the direction of the top spin analyzer (non-$b$ jet) 
in the top rest frame and the top spin quantization axis ($s_t$) in the 
$t\bar{t}$-ZMF. The angle $\cos\theta_{qs_t}$ is the angle $\cos\theta_f$ defined
in Eq.~(\ref{eq:diffsigma}). The spin of the top can be chosen in the direction
of any of the spin quantization axes as defined in Sec.~\ref{sec:spintop}. 
This distribution is sensitive to the polarization of the top and we show in 
Fig.~\ref{fig:basis} the distribution calculated in the different bases. As 
expected, the `beamline' basis and the `off-diagonal' basis are most sensitive 
to the top polarization and therefore also to the decay dynamics of the top.
The chiral nature of the FCNC coupling will be more clearly visible in these 
two basis, with a flat distribution in case of the equality of the two chiral 
coupling. The `helicity' and the `minimal' basis will not be effective in 
discriminating the chirality and they are shown just for the illustration. 
The effect is further enhanced with the  
beam polarizations of $P^L_{e^-} = -0.8$ and $P^L_{e^+} = 0.3$,
in all the spin bases considered here. We show the distribution in the `off-diagonal' basis 
in Fig.~\ref{fig:offDB}, as it is most sensitive to the beam polarization.

The double differential angular distribution of the top and the antitop
defined in Eq.~(\ref{eq:diffsigma}) provides a measurement of the spin-spin 
correlations. It was shown in Ref.~\cite{Bernreuther:2013aga} that, for the  
experimental analysis, it is more suitable to use the one-dimensional 
distribution of the product of the cosines, ${\cal O}_{s_t,s_{\bar{t}}} =
\cos \theta_f \cos \theta_{\bar{f}}$, rather than analyzing 
Eq.~(\ref{eq:diffsigma}). We define $\cos \theta_f \cos \theta_{\bar{f}}$ 
as $\cos\theta_{qs_t} \cos\theta_{ls_{\bar{t}}}$ for our analysis. The 
$1/\sigma (d\sigma/d{\cal O}_{s_t,s_{\bar{t}}})$ distribution is shown 
in Fig.~\ref{fig:coscos}, using the `off-diagonal' basis and a longitudinal
beam polarization of  $P^L_{e^-} = -0.8$ and $P^L_{e^+} = 0.3$ . 
The asymmetry of the plot around $\cos\theta_{qs_t} \cos\theta_{ls_{\bar{t}}}$ = 0, 
signals for the spin-spin correlation. The plot for Case 2 ($|g_{qt}|^2$ = 0) shows more events for 
positive values for $\cos\theta_{qs_t} \cos\theta_{ls_{\bar{t}}}$, 
whereas for Case 3 ($|g_{tq}|^2$ = 0)
one gets more events for negative values of $\cos\theta_{qs_t} \cos\theta_{ls_{\bar{t}}}$. 

\begin{figure}
\centering
\begin{minipage}{0.45\textwidth}
\centering
\includegraphics[width=7cm, height=6cm]{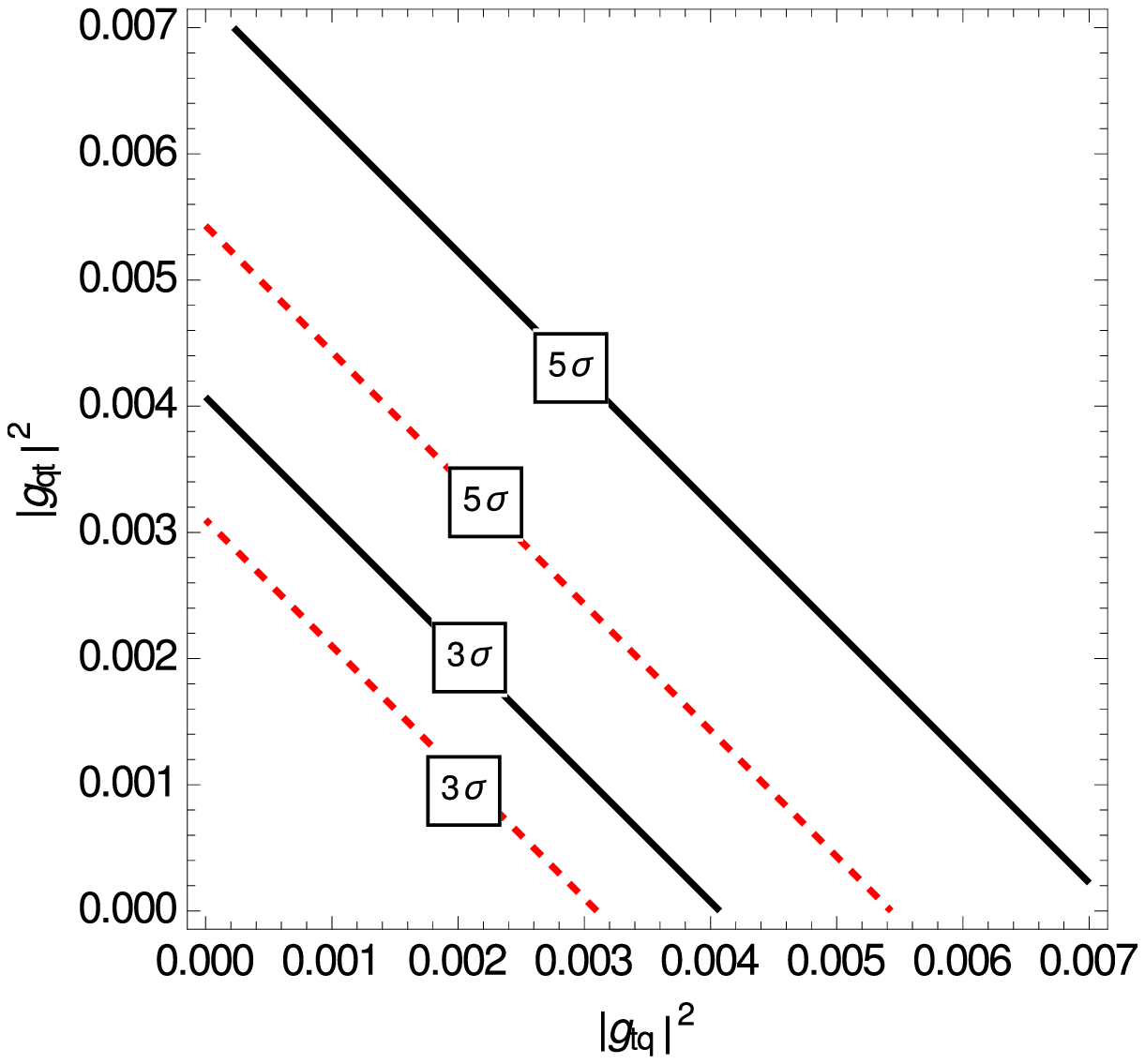}
\caption{Contour plots in the $|g_{tq}|^2-|g_{qt}|^2$ plane, for the statistical significance 
$S$, from the production cross section, at $\sqrt{s}$ = 500 GeV and a luminosity of 500 fb$^{-1}$, with
unpolarized beams [black] and a beam polarization of $P^L_{e^-}$ = -0.8 and $P^L_{e^+}$ = 0.3 [red-dashed].}
\label{fig:sig1}
\end{minipage}
\hspace{0.2cm}
\begin{minipage}{0.45\textwidth}
\centering
\includegraphics[width=7cm, height=6cm]{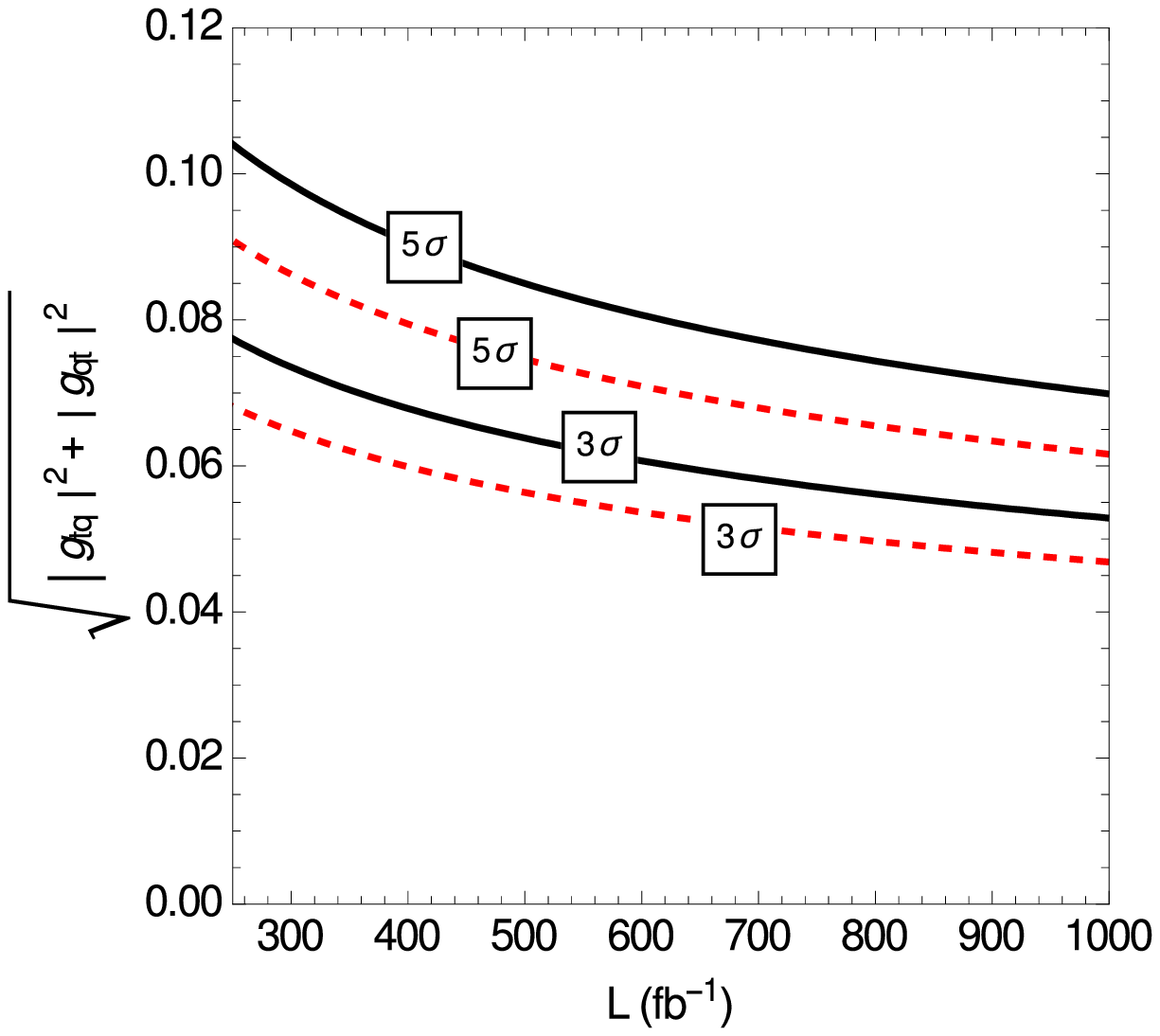}
\caption{The sensitivity of 3$\sigma$ and 5$\sigma$ to the FCNC coupling $\sqrt{|g_{tq}|^2 +|g_{qt}|^2}$
at $\sqrt{s}$ = 500 GeV, as a function of integrated luminosity. The black solid line is for unpolarized beams,
and the red-dashed line is for a beam polarization of $P^L_{e^-}$ = -0.8 and $P^L_{e^+}$ = 0.3. }
\label{fig:sig2}
\end{minipage}
\end{figure}

We next estimate the sensitivity that can be obtained for the FCNC $tqH$ couplings, 
given by the efficient signal identification and the significant background suppression
which can be achieved at the linear collider. We adopt the following formula for 
the significance measurement~\cite{Cowan:2010js},
\begin{equation}
 S = \sqrt{2\left[\left(N_S+N_B\right) {\rm ln}\left(1+\frac{N_S}{N_B}\right)-N_S\right]},
\end{equation}
with $N_S$ and $N_B$ being the number of signal and background events. 
In Fig.~\ref{fig:sig1} we present the contours of 3$\sigma$ and 5$\sigma$ significance
for our process in the $|g_{tq}|^2-|g_{qt}|^2$ plane. The sensitivity of the linear 
collider will increase with the implementation of beam polarization with left polarized 
electrons and right polarized positrons. Since the total cross section is  proportional
to $|g_{tq}|^2+|g_{qt}|^2$, the contours are symmetric in that plane. 
The sensitivity to the coupling $\sqrt{|g_{tq}|^2+|g_{qt}|^2}$, as a function of the 
integrated luminosity for $\sqrt{s}$ = 500 GeV is shown Fig.~\ref{fig:sig2}. 
One can see that at 3$\sigma$ statistical sensitivity and 
${\cal L}$ = 500 fb$^{-1}$, $\sqrt{|g_{tq}|^2+|g_{qt}|^2}$
can be probed to 0.063 (0.056) with unpolarized (polarized) beams. The limits obtained from the asymmetries, specially 
$A_{fb}$ from Sec.~\ref{sec:cmf_asymmetries} will be more stronger
and will not be symmetric in the $|g_{tq}|^2-|g_{qt}|^2$ plane.
We find the following upper bounds as listed in Table~\ref{tab:limits} at the 2$\sigma$, 
3$\sigma$ and the 5$\sigma$ level from the total cross section, in the case of the 
polarized and the unpolarized beams.
\begin{table}[htb]
\begin{center}
\begin{tabular}{|c|c|c|c|c|} \hline
 &\multicolumn{2}{|c|}{$P^L_{e^-} = 0, P^L_{e^+} = 0$}&\multicolumn{2}{|c|}{$P^L_{e^-} = -0.8, P^L_{e^+} = 0.3$} \\ \hline
 &&&& \\
Significance &$\sqrt{|g_{tq}|^2+|g_{qt}|^2}$ &BR($t\rightarrow qH$)
&$\sqrt{|g_{tq}|^2+|g_{qt}|^2}$ &BR($t\rightarrow qH$) \\ \hline
2$\sigma$ &0.052                &7.61$\times 10^{-4}$               &0.046                &5.96$\times 10^{-4}$  \\ \hline
3$\sigma$ &0.063                &1.19$\times 10^{-3}$                &0.056               &8.84$\times 10^{-4}$   \\ \hline
5$\sigma$ &0.085                &2.04$\times 10^{-3}$                &0.074               &1.54$\times 10^{-3}$   \\ \hline
\end{tabular}
\caption{Upper bounds on $\sqrt{|g_{tq}|^2+|g_{qt}|^2}$ and the respective branching ratios,
that can be obtained in the ILC, at $\sqrt{s}$ = 500 GeV, with a luminosity of 500 fb$^{-1}$.
The results are presented for both, the polarized and the unpolarized case.}
\label{tab:limits}
\end{center}
\end{table}

\section{Conclusion}\label{sec:conclusion}

We have studied the flavor violating top-Higgs interactions, 
at the $e^-e^+$ linear colliders using different beam polarizations.  There are 
several works exhibiting the prospects of the LHC to constrain or discover 
these couplings, by considering several signatures of the flavor violating interactions.
The LHC experiments have also looked into these couplings and have obtained bounds on the branching ratio
of the process $t\rightarrow q H$. These flavor violating interactions can have a chiral structure with the top coupling differently to the left handed and the right handed fermions. 
Since the branching ratio of the top to $qH$, as well as, the total 
production cross section is being proportional to $|g_{tq}|^2+|g_{qt}|^2$, the chiral
nature won't be evident from these measurements.

Therefore, we have looked in the context of the linear collider into various observables  
which will highlight this aspect of the couplings. The polar angle distribution
of the quark emitted from the $t\rightarrow cH$ decay, will exhibit a behaviour
sensitive to the nature of the coupling. This will change with
the change of the beam polarization.  The distribution will be flat for all the 
polarization combinations if $|g_{tq}|^2 = |g_{qt}|^2$. The presence of only
one of the coupling ($|g_{tq}|^2$) leads to a forward peak for $e^-_L e^+_R$ polarization
and will be unchanged for the $e^-_R e^+_L$ polarization. The opposite behaviour is observed for 
$|g_{qt}|^2$. Next, the forward-backward asymmetry $A_{fb}$ is used in order
to constrain the $|g_{tq}|^2-|g_{qt}|^2$ parameter space. 

The spins of the tops are correlated in the top pair production and the decay products of the 
tops are correlated with the spins, therefore the decay products of the top and the antitop
are correlated. 
%The angular distribution of the top decay products being correlated with
%the top spin axis, 
The presence of new physics in the top decay will therefore, lead to a change in 
the correlation coefficient in the angular distribution of the top decay products. A right 
choice of spin basis of the top quark pair
is also important in enhancing the correlation. We consider different 
observables in Sec.~\ref{sec:spintop}, which are sensitive to the spin analyzing power
($\kappa$) of the top decay product. The quark emitted
from the top FCNC decay, will be a perfect spin analyzer ($\kappa_q =1$)
in the presence of a single chiral coupling. The $\kappa_q$ of the emitted quark 
will be zero when $|g_{tq}|^2 = |g_{qt}|^2$ and the correlation will be lost. 
We have performed an analysis applying
all the cuts at the linear collider in Sec.~\ref{sec:numericalstudy}, and have studied the spin observables
in the context of different spin bases. We find that the off-diagonal basis along with 
the beamline basis are the most sensitive to the chirality of the couplings. The  effect is even more enhanced by polarizing 
the initial beams of left handed electrons and right handed positrons.

Finally, we have obtained a limit on the couplings from the total cross section and find 
that BR($t\rightarrow qH$) can be probed to $5.59 \times 10^{-3}(8.84 \times 10^{-4})$
at 3$\sigma$ level at the ILC, with $\sqrt{s} =$ 500 GeV, $\cal L$ = 500 fb$^{-1}$ and 
a beam polarization of $P^L_{e^-} = 0 (-0.8), P^L_{e^+} = 0 (0.3)$ , which hopefully will be observed 
at the future linear colliders. 

\section*{Acknowledgments}
We would like to thank Juan Antonio Aguilar Saavedra for very useful discussions. 
This work is supported by the Croatian Science Foundation (HRZZ) project PhySMaB,
``Physics of Standard Model and Beyond" as well as by the H2020 Twinning
project No. 692194, ``RBI-T-WINNING''.

\appendix
\section{Helicity amplitudes for the production and the decay}\label{sec:All_appen}

\subsection{The production $e^-e^+ \rightarrow t\bar{t}$}\label{sec:Appendix1}
The helicity amplitudes for the process $e^-e^+ \rightarrow t\bar{t}$ are defined below. They are the 
same as those considered in~\cite{Grzadkowski:1995te}, with the normalization factor taken care of.  The amplitudes
%of $e^-$, $e^+$, $t$ and $\bar{t}$ 
are defined as $\mathcal{M}_{LRIJ}$, where $L$ denotes the 
left-handed electron beam $e^-_L$, $R$ for right-handed positron beam $e^+_R$, and $IJ$ denotes the different 
final-state combinations of $t\bar{t}$, i.e. $\downarrow\downarrow$, $\downarrow\uparrow$, 
$\uparrow\downarrow$ and $\uparrow\uparrow$.  Similarly $\mathcal{M}_{RLIJ}$ 
denotes the right-handed electron beam $e^-_R$ and left-handed
positron beam $e^+_L$. For the helicity-conserving interactions, the amplitudes are as follows:
\begin{eqnarray}\label{va_hel}
&&\mathcal{M}_{LR\uparrow\uparrow} =B A_L m_t \sin \theta_t  \,,\\ \nonumber
&&\mathcal{M}_{LR\uparrow\downarrow} =B (E A_L + k B_L) (1 + \cos \theta_t) \,,\\ \nonumber
&&\mathcal{M}_{LR\downarrow\uparrow} =-B (E A_L - k B_L) (1 - \cos \theta_t) \,, \\ \nonumber
&&\mathcal{M}_{LR\downarrow\downarrow} =-B A_L m_t \sin \theta_t \,,\\ \nonumber
&&\mathcal{M}_{RL\uparrow\uparrow} =B A_R m_t \sin \theta_t  \,, \\ \nonumber
&&\mathcal{M}_{RL\uparrow\downarrow} =-B (E A_R + k B_R) (1 - \cos \theta_t) \,, \\ \nonumber
&&\mathcal{M}_{RL\downarrow\uparrow} =B (E A_R - k B_R) (1 + \cos \theta_t) \,,  \\ \nonumber
&&\mathcal{M}_{RL\downarrow\downarrow} =-B A_R m_t \sin \theta_t.   
\end{eqnarray}
All the expressions above have the normalization factor $B$ defined as  $i\sqrt{3 \beta \alpha^2/4}$.
$E$ is the  beam  energy $\sqrt{s}/2$ and $k = E\beta$, where $\beta =\sqrt{1-4m_t^2/s}$. The other constants which appear are defined below:
\begin{eqnarray}
A_L&=&\frac{2}{s}Q_t Q_e+\frac{2g_t^V}{s-mZ^2}(g_e^V+g_e^A),~~A_R
=\frac{2}{s}Q_t Q_e+\frac{2g_t^V}{s-mZ^2}(g_e^V-g_e^A) \,, \nonumber \\
B_L&=&\frac{2g_t^A}{s-mZ^2}(g_e^V+g_e^A),~~B_R=\frac{2g_t^A}{s-mZ^2}(-g_e^V+g_e^A),
%\nonumber \\ B&=&\frac{\alpha}{2}\sqrt{\frac{3\beta}{s}}
\label{eq:albl}
\end{eqnarray}
where $Q_e = -1,~Q_t=2/3$, $\theta_W$ is the Weinberg mixing angle and
\begin{eqnarray}\label{eq:geVgeA}
g_e^V &=& \frac{e}{\sin 2\theta_W}\left(-\frac{1}{2} + 2 \sin^2\theta_W\right),~~~~
g_e^A = -\frac{e}{2 \sin 2\theta_W} \,, \nonumber \\
g_t^V &=& \frac{e}{\sin 2\theta_W}\left(\frac{1}{2} - \frac{4}{3} \sin^2\theta_W\right),~~~~
g_t^A = \frac{e}{2 \sin 2\theta_W}.
\end{eqnarray}

\subsection{The decay $t\rightarrow q H$} \label{sec:appen_decaytcH}
The squared matrix elements $\rho^{D(t)}_{\lambda_t\lambda_t'}$ for the top quark
decay in its rest frame is given by
\begin{eqnarray}\label{eq_me_topch_rest}
\rho^{D(t)}_{\uparrow \uparrow}&=& m_t\left[E^{top}_q\left\lbrace |g_{tq}|^2\left(1-\cos\theta^{top}_q\right)+
|g_{qt}|^2\left(1+\cos\theta^{top}_q\right)\right\rbrace+2 m_q g_{tq} g_{qt}\right]\,, \\ \nonumber
\rho^{D(t)}_{\downarrow \downarrow}&=& m_t\left[E^{top}_q\left\lbrace |g_{tq}|^2\left(1+\cos\theta^{top}_q\right)+
|g_{qt}|^2\left(1-\cos\theta^{top}_q\right)\right\rbrace+2 m_q g_{tq} g_{qt}\right]\,, \\ \nonumber
 \rho^{D(t)}_{\uparrow \downarrow}&=& -E^{top}_q m_t \sin \theta^{top}_q e^{i\phi^{top}_{q}} (|g_{tq}|^2-|g_{qt}|^2) \,, \\ \nonumber
\rho^{D(t)}_{\downarrow \uparrow}&=& -E^{top}_q m_t \sin \theta^{top}_q e^{-i\phi^{top}_{q}} (|g_{tq}|^2-|g_{qt}|^2),
 \end{eqnarray}
where $E_q^{top},~\theta_q^{top},~\phi_q^{top}$ are the energy and the polar and 
the azimuthal angle of the emitted quark $q$ in the top rest frame, respectively. We obtain
the relevant $\rho^{D(t)}_{\lambda_t\lambda_t'}$ in the c.m. frame 
by making the following substitution in the above equations:
\begin{eqnarray}\label{eq:boost}
 E^{top}_q &=&E_q \left( \frac{\sqrt{1-\beta^2}}{1+\beta \cos\theta_q}\right), \nonumber \\
 \cos\theta^{top}_q &=&\frac{\beta-\cos\theta_{tq}}{\beta\cos\theta_{tq}-1}, \nonumber \\
 \sin \theta^{top}_q e^{\pm i \phi^{top}_q}&=&\frac{\sqrt{1-\beta^2}}{1-\beta\cos\theta_{tq}}
 \left(\cos \theta_t \sin \theta_q \cos \phi_q - \sin \theta_t \cos \theta_q \pm i \sin \theta_q \sin \phi_q\right),
\end{eqnarray}
where $\theta_{tq}$ and $E_q$ are defined in Eq.~(\ref{eq:cosTHtq}).
The squared matrix elements is similar for antitop, with $\beta$ replaced
by $-\beta$. We have assumed $m_q$ = 0, for all our calculations.

\subsection{The decays $t\rightarrow W b$ and $t\rightarrow l^+ \nu b$} \label{sec:appen_decaytWb}
The squared matrix elements $\rho^{D(t)}_{\lambda_t\lambda_t'}$ for the top quark 
decaying to $W^+b$, in the rest frame of the top is given by
\begin{eqnarray}\label{eq_me_topWB_rest}
\rho^{D(t)}_{\uparrow \uparrow}&=& \frac{E^{top}_q g^2 m_t}{2 m_W^2}  \left[
(1- \cos \theta^{top}_{q})m_t^2+2 m_W^2 (1+ \cos \theta^{top}_{q})\right] \,, \\ \nonumber
\rho^{D(t)}_{\downarrow \downarrow}&=&\frac{E^{top}_q g^2 m_t}{2 m_W^2}  \left[
(1+\cos \theta^{top}_{q})m_t^2+2 m_W^2 (1- \cos \theta^{top}_{q})\right] \,, \\ \nonumber
 \rho^{D(t)}_{\uparrow \downarrow}&=& \frac{E^{top}_q g^2 m_t}{2 m_W^2}  (m_t^2-2 m_W^2) \sin\theta^{top}_q
 e^{i\phi^{top}_{q}} \,, \\ \nonumber
\rho^{D(t)}_{\downarrow \uparrow}&=&  \frac{E^{top}_q g^2 m_t}{2 m_W^2}  (m_t^2-2 m_W^2) \sin\theta^{top}_q
e^{-i\phi^{top}_{q}}\,. 
 \end{eqnarray}
The squared matrix elements $\rho^{D(t)}_{\lambda_t\lambda_t'}$ for the top quark decay to 
 $l^+ \nu b$ in its rest frame, is 
\begin{eqnarray}\label{eq_me_top}
 \rho^{D(t)}_{\uparrow \uparrow}&=& \frac{g^4 E_l m_t (m_t^2-m_b^2-2 p_t \cdot p_l)}
 {(p_W^2-M_W^2)^2+M_W^2 \Gamma_W^2} (1+\cos \theta^{top}_{l^+}) \,, \\ \nonumber
\rho^{D(t)}_{\downarrow \downarrow}&=& \frac{g^4 E_l m_t (m_t^2-m_b^2-2 p_t\cdot p_l)}
 {(p_W^2-M_W^2)^2+M_W^2 \Gamma_W^2} (1-\cos \theta^{top}_{l^+}) \,, \\ \nonumber
 \rho^{D(t)}_{\uparrow \downarrow}&=& \frac{g^4 E_l m_t (m_t^2-m_b^2-2 p_t\cdot p_l)}
 {(p_W^2-M_W^2)^2+M_W^2 \Gamma_W^2} (\sin \theta^{top}_l e^{i\phi^{top}_{l^+}}) \,, \\ \nonumber
\rho^{D(t)}_{\downarrow \uparrow}&=& \frac{g^4 E_l m_t (m_t^2-m_b^2-2 p_t\cdot p_l)}
 {(p_W^2-M_W^2)^2+M_W^2 \Gamma_W^2} (\sin \theta^{top}_l e^{-i\phi^{top}_{l^+}}) ,
 \end{eqnarray}
where $E_f^{top},~\theta_f^{top},~\phi_f^{top}$ are the energy, and the polar and the azimuthal
angle of the emitted fermion in the top rest frame, respectively. The emitted fermion is $b$ 
in Eq.~(\ref{eq_me_topWB_rest}) from the decay $t\rightarrow W^+b$ and 
$l$ from the decay $t\rightarrow l^+\nu b$ in Eq.~(\ref{eq_me_top}).
The squared matrix elements $\rho^{D(t)}_{\lambda_t\lambda_t'}$ for the top quark 
decaying to $W^+b$ ($l^+ \nu b$) in the c.m frame is obtained by inserting 
the substitutions from Eq.~(\ref{eq:boost}), to Eqs.~(\ref{eq_me_topWB_rest}) 
and (\ref{eq_me_top}). The squared matrix elements for the antitop is obtained 
by replacing $\beta$ with $-\beta$.
 
\section{Observables (${\cal O}_{1,2,3,4})$} \label{sec:appen_spin}

Below we list the analytical expressions for the various spin observables calculated and considered in Sec.~\ref{sec:spintop}. 
\begin{align}
{\cal O}_1 &= \frac{1}{3\sigma_{t\bar{t}}}(1-P^L_{e^-}P^L_{e^+}) \left[(3-\beta^2) \left\lbrace C^2_\gamma +
 \frac{2s}{s-m_Z^2} C_\gamma C_Z g_t^V \left(g_e^V+ P_{eff} g_e^A \right)\right\rbrace\right. \nonumber \\
 &\left. + \frac{s^2}{(s-m_Z^2)^2} C_Z^2\left((3-\beta^2) (g_t^{V})^2 +2 \beta^2 (g_t^A)^2\right)
\left((g_e^V)^2+ (g_e^A)^2+2 P_{eff} g_e^V g_e^A \right) \right] \label{eq:O1}\\
{\cal O}^{hel}_2 &= -\frac{2\beta}{3\sigma_{t\bar{t}}} (1-P^L_{e^-}P^L_{e^+})\left[\frac{s}{s-m_Z^2}C_\gamma
C_Z g_t^A \left(g_e^V+P_{eff} g_e^A\right) \right. \nonumber \\
&\left. + \frac{s^2}{(s-m_Z^2)^2} C_Z^2 g_t^A g_t^V
\left((g_e^V)^2+ (g_e^A)^2+2 P_{eff} g_e^V g_e^A \right) \right] \label{eq:O2h}\\
{\cal O}^{beam}_2 &= -\frac{1}{6\sigma_{t\bar{t}}}(1-P^L_{e^-}P^L_{e^+})\left[(2\sqrt{1-\beta^2}+1)
\left\lbrace C^2_\gamma P_{eff} +\frac{2s}{s-m_Z^2}C_\gamma C_Z g_t^V
\left(g_e^A+ P_{eff} g_e^V \right) \right\rbrace \right. \nonumber \\
 &\left. + \frac{s^2}{(s-m_Z^2)^2} C_Z^2\left((2\sqrt{1-\beta^2}+1)(g_t^V)^2 
 +\beta^2 (g_t^A)^2\right) \left( P_{eff}((g_e^V)^2+ (g_e^A)^2)+2 g_e^V g_e^A \right) \right] \label{eq:O2b}\\
{\cal O}^{off}_2 &= -\frac{1}{4\beta\sigma_{t\bar{t}}}(1-P^L_{e^-}P^L_{e^+})\left[
 ((\beta^2-1) \tanh^{-1}\beta-\beta) \right. \nonumber \\
&\left. \left\lbrace C^2_\gamma P_{eff} +\frac{2s}{s-m_Z^2}C_\gamma C_Z g_t^V
\left(g_e^A+ P_{eff} g_e^V \right) \right\rbrace 
  + \frac{s^2 \chi}{(s-m_Z^2)^2} C_Z^2  \right. \nonumber \\
&\left. \left((\beta^2-1) \tanh^{-1}\beta
 -\beta \frac{(g_t^V)^2+(g_t^A)^2}{(g_t^V)^2- (g_t^A)^2}
 \right) \left( P_{eff}\left((g_e^V)^2+ (g_e^A)^2\right)+2 g_e^V g_e^A \right) \right] \label{eq:O2o}\\
 {\cal O}^{min}_2 &= -\frac{1}{2\sigma_{t\bar{t}}}\sqrt{\beta^2-1}(1-P^L_{e^-}P^L_{e^+})
 \left.E\left[\frac{\sin^{-1}\left(\frac{\beta}{\sqrt{\beta^2-1}}\right)}
 {\sqrt{\beta+1}} \right| 1-\frac{1}{\beta^2}\right]
 \left[ \frac{s}{s-m_Z^2}C_\gamma C_Z g_t^A  \right. \nonumber \\
 &\left. \left(g_e^V+ P_{eff} g_e^A \right) + \frac{s^2}{(s-m_Z^2)^2} C_Z^2 \beta g_t^A g_t^V 
 \left( (g_e^V)^2+ (g_e^A)^2+2 P_{eff} g_e^V g_e^A \right) \right] \label{eq:O2m}\\ 
  {\cal O}^{hel}_3 &=-\frac{1}{3\sigma_{t\bar{t}}}(1-P^L_{e^-}P^L_{e^+})\left[C^2_\gamma (1+\beta^2) +
 \frac{2s}{s-m_Z^2} C_\gamma C_Z g_t^V \left(g_e^V+ P_{eff} g_e^A \right)(1+\beta^2)\right. \nonumber \\
 &\left. + \frac{s^2}{(s-m_Z^2)^2} C_Z^2\left((1+\beta^2)(g_t^V)^2 +2 \beta^2 (g_t^A)^2\right)
\left((g_e^V)^2+ (g_e^A)^2+2 P_{eff} g_e^V g_e^A \right) \right] \label{eq:O3h}\\
{\cal O}^{beam}_3 &=\frac{1}{15\sigma_{t\bar{t}}}(1-P^L_{e^-}P^L_{e^+})
(-3\beta^2+4\sqrt{1-\beta^2}+11)\left[C^2_\gamma +
 \frac{2s}{s-m_Z^2} C_\gamma C_Z g_t^V \right. \nonumber \\
 &\left. \left(g_e^V+ P_{eff} g_e^A \right) + \frac{s^2}{(s-m_Z^2)^2} C_Z^2 (g_t^V)^2 
\left((g_e^V)^2+ (g_e^A)^2+2 P_{eff} g_e^V g_e^A \right) \right] \label{eq:O3b}\\
{\cal O}^{off}_3 &=\frac{1}{3\sigma_{t\bar{t}}}(1-P^L_{e^-}P^L_{e^+})\left[(3-\beta^2)
\left\lbrace C^2_\gamma + 
 \frac{2s}{s-m_Z^2} C_\gamma C_Z g_t^V \left(g_e^V+ P_{eff} g_e^A \right)\right\rbrace \right. \nonumber \\
 &\left. - \frac{s^2}{\beta(s-m_Z^2)^2} C_Z^2
 \left((g_e^V)^2+ (g_e^A)^2+2 P_{eff} g_e^V g_e^A \right) \right. \nonumber \\ 
 &\left. \left(3\sqrt{1-\beta^2} \tan^{-1}\left(\frac{\beta}{\sqrt{1-\beta^2}}\right) (g_t^A)^2 -
 \beta(3- \beta^2)((g_t^V)^2+ (g_t^A)^2)\right) \right] \label{eq:O3o}\\
{\cal O}^{min}_3 &=\frac{1}{3\sigma_{t\bar{t}}}(1-P^L_{e^-}P^L_{e^+})\left[\beta^2
\left\lbrace C^2_\gamma +  \frac{2s}{s-m_Z^2} C_\gamma C_Z g_t^V \left(g_e^V+ P_{eff} g_e^A \right)\right\rbrace \right. \nonumber \\
 &\left. + \frac{s^2}{\beta(s-m_Z^2)^2} C_Z^2
 \left((g_e^V)^2+ (g_e^A)^2+2 P_{eff} g_e^V g_e^A \right) \right. \nonumber \\ 
 &\left. \left(3\sqrt{1-\beta^2} \tan^{-1}\left(\frac{\beta}{\sqrt{1-\beta^2}}\right) (g_t^A)^2
 +\beta^3 (g_t^V)^2 + \beta(2\beta^2-3) (g_t^A)^2\right) \right] \label{eq:O3m}\\
{\cal O}_4 &= \frac{4\beta}{3\sigma_{t\bar{t}}} (1-P^L_{e^-}P^L_{e^+})\left[\frac{s}{s-m_Z^2}C_\gamma
C_Z g_t^A \left(g_e^A+P_{eff} g_e^V\right) \right. \nonumber \\
&\left. + \frac{s^2}{(s-m_Z^2)^2} C_Z^2 g_t^A g_t^V
\left(P_{eff}((g_e^V)^2+ (g_e^A)^2)+2  g_e^V g_e^A \right) \right]\label{eq:O4},
\end{align}
where $\sigma_{t\bar{t}}$ is the total cross section given by
\begin{align}
\sigma_{t\bar{t}} &= \frac{1}{3}(1-P^L_{e^-}P^L_{e^+}) \left[(3-\beta^2) 
\left\lbrace C_\gamma^2+\frac{2s}{s-m_Z^2} C_\gamma C_Z g_t^V \left(g_e^V+ P_{eff} g_e^A \right)\right\rbrace\right. \nonumber \\
&\left. + \frac{s^2}{(s-m_Z^2)^2} C_Z^2\left((3-\beta^2) (g_t^V)^2 + 2\beta^2 (g_t^A)^2\right)
\left((g_e^V)^2+ (g_e^A)^2+2 P_{eff} g_e^V g_e^A \right) \right] \label{eq:SMTT},
\end{align}
and $P_{eff}=(P^L_{e^-}-P^L_{e^+})/(1-P^L_{e^-}P^L_{e^+}),~C_\gamma=e^2 Q_t Q_e,~C_Z=g^2/(2\cos\theta_W)^2,
~\chi= ((g_t^V)^2- (g_t^A)^2)$. The terms $g_{t,e}^{A,V}$ are given in Eq.~(\ref{eq:geVgeA}) and $E[..|..]$
%${\cal E}_l$ 
in Eq.~(\ref{eq:O2m}) is the elliptic function of the second kind.
% and is given by
%\begin{equation*}
% \left.E\left[\frac{\sin^{-1}\left(\frac{\beta}{\sqrt{\beta^2-1}}\right)}
% {\sqrt{\beta+1}} \right| 1-\frac{1}{\beta^2}\right].
%\end{equation*}

\end{document}